\documentclass[12pt]{article}%
\usepackage[nosort]{cite}
\usepackage{graphicx}
\usepackage{multicol}
\usepackage{amsfonts}
\usepackage{amssymb}
\usepackage{amsmath}
\usepackage{heck}
\usepackage[all]{xy}
\xyoption{arc}
\usepackage{setspace}
\usepackage{verbatim}
\usepackage{color}
\usepackage{epsfig}
\usepackage[margin=1in]{geometry}
\usepackage{titletoc}
\usepackage{lscape}
\usepackage{tikz}
\usepackage[debug,pageanchor=false]{hyperref}%
\providecommand{\U}[1]{\protect\rule{.1in}{.1in}}
\numberwithin{equation}{section}
\def\Z{\mathbb{Z}}

\def\C{\mathbb{C}}

\renewcommand{\a}{\alpha}
\renewcommand{\b}{\beta}

\newcommand{\cN}{\mathcal{N}}
\newcommand{\cO}{\mathcal{O}}
\newcommand{\cS}{\mathcal{S}}

\newcommand{\cT}{\mathcal{T}}

\newcommand\PP{\mathbb{P}}

\newcommand{\be}{\begin{equation}}
\newcommand{\ee}{\end{equation}}

\definecolor{link}{rgb}{0,0,0}
\hypersetup{colorlinks=true,linkcolor=link,citecolor=link,urlcolor=link,linktocpage}
\begin{document}

\date{April 2015}

\title{Geometric Engineering, \\ \vspace{-1.2cm} Mirror Symmetry \\ \vspace{-1.2cm} and \\ $6d_{(1,0)} \rightarrow 4d_{({\cal N}=2)}$}

\institution{HARVARD}{\centerline{${}^{1}$ Jefferson Physical Laboratory, Harvard University, Cambridge, MA 02138, USA}}
\institution{HARVARDmath}{\centerline{${}^{2}$ CMSA, Harvard University, Cambridge, MA 02138, USA}}

\authors{Michele Del Zotto\worksat{\HARVARD}\footnote{e-mail: {\tt delzotto@physics.harvard.edu}},
Cumrun Vafa\worksat{\HARVARD}\footnote{e-mail: {\tt vafa@physics.harvard.edu}}, and Dan Xie\worksat{\HARVARD,\HARVARDmath}\footnote{e-mail: {\tt dxie@cmsa.fas.harvard.edu}}}

\abstract{We study compactification of 6 dimensional (1,0) theories on $T^2$. We use geometric engineering
of these theories via F-theory and employ mirror symmetry technology to solve for the effective 4d ${\cal N}=2$ geometry
for a large number of the $(1,0)$ theories including those associated with conformal matter.
Using this we show that for a given 6d theory we can obtain many inequivalent 4d ${\cal N}=2$ SCFTs.
Some of these respect the global symmetries of the 6d theory while others exhibit $SL(2,\Z)$ duality
symmetry inherited from global diffeomorphisms of the $T^2$.  This construction also explains the 6d origin of moduli space of 4d affine $ADE$ quiver theories
as flat $ADE$ connections on $T^2$.   Among
the resulting $4d$ ${\cal N}=2$ CFTs we find theories
whose vacuum geometry is captured by an LG theory (as opposed to a curve or a local CY geometry). 
We obtain arbitrary genus curves of class ${\cal S}$ with punctures from toroidal compactification of $(1,0)$ SCFTs
where the curve of the class ${\cal S}$ theory emerges through mirror symmetry.
We also show that toroidal compactification of the little string version of these theories can lead to
class ${\cal S}$ theories with no punctures on arbitrary genus Riemann surface. }

\maketitle

\tableofcontents

\enlargethispage{\baselineskip}

\setcounter{tocdepth}{2}

\section{Introduction \label{sec:INTRO}}
Nontrivial properties of a lower dimensional quantum field theory could be made manifest if they can be derived from the compactification of a higher dimensional theory. The typical
example is four dimensional $\mathcal{N}=4$ SYM whose $SL(2,\Z)$ duality is best understood by using the $T^2$ compactification 
of 6d $(2,0)$ theory \cite{Witten:1995zh}. Similarly, the S duality of four dimensional  $\mathcal{N}=2$ class ${\cal S}$ theories could be derived from 
compactifying 6d $(2,0)$ theory on a punctured Riemann surface \cite{Gaiotto:2009we}. 

Recently, a classification of 6d (1,0) theories has been proposed which is a surprisingly rich set\cite{Heckman:2013pva,Heckman:2015bfa} (see also \cite{Bhardwaj:2015xxa}). It is natural to ask what kind of 4d theory we can get and what  kind of 
interesting 4d dynamics we can learn from their compactification.  
In principle we can get ${\cal N}=1$ or ${\cal N}=2$ theories in 4d.  
The simplest case to start with would be the ${\cal N}=2$ which arises by considering $T^2$ compactification.
Such compactification has been studied for E-string theory \cite{Klemm:1996hh,Ganor:1996pc} and recently for  6d minimal conformal matter \cite{Ohmori:2015pua}. 

The purpose of this work is to study $T^2$ compactification for a broader class of 6d $(1,0)$ SCFTs and see
what lessons one learns.  Naively, one may expect not too many new discoveries as 
we can only use the torus to do the compactification. However, our study shows that the story is surprisingly interesting and rich. 
A large class of examples arise from studying 6d SCFTs which can be geometrically engineered by orbifolds in F-theory.  We then
use the duality with type IIA upon $T^2$ compactification and
mirror symmetry for $(T^2\times \C^2)/\mathsf{G}$ orbifolds \cite{Hori:2000kt,Vafa:2001ra} to obtain the effective
4d ${\cal N}=2$ geometry (which is typically a local Calabi-Yau 3-fold). Using this we write down the full effective 4d ${\cal N}=2$ geometry for 
the 6d theories on $T^2$. 

To find interesting conformal theories in 4d we try to locate a maximal singular point from our ${\cal N}=2$ geometry.
It turns out that there are two roads to locate a four dimensional $\mathcal{N}=2$ SCFT. If we keep the complex structure of the torus $\tau$ as the exact marginal 
deformation, we get a 4d gauge theory whose gauge coupling is identified with $\tau$ and has a natural $SL(2,\Z)$ duality symmetry. 
Therefore we find a large class of new 4d $\mathcal{N}=2$ theory with $SL(2,\Z)$ duality group, which are the generalizations of
the 6d $(2,0)$ origin of $SL(2,\Z)$ duality symmetry for the 4d $\mathcal{N}=4$ SYM.  Just as in the ${\cal N}=4$ case, these
are the cases where compactification to 5d do not yield a conformal theory but to 4d does, so the CFT skips a dimension
and goes from 6 directly to 4.  The 4d affine $ADE$ quiver theories of \cite{Witten:1997sc,Katz:1997eq} is in this class.  They arise
in 6d theories in which $ADE$ is part of the global symmetry.  Turning on Wilson lines for these global symmetries on $T^2$ leads
to moduli for the 4d theory.  Moreover, this provides a 6d explanation for the identification of the moduli space of the resulting 4d theory as the
space of flat $ADE$ connections on $T^2$.  A large number of these theories are realized by considering
F-theory on
orbifold elliptic 3-folds which we study in detail.   From these orbifold theories we also obtain 4d SCFTs which are $A,D$ and $E$ gauge theories where the matter involves gauging three or four copies of $D_p(G=ADE)$ \cite{Xie:2012hs, Cecotti:2012jx,Cecotti:2013lda} (which are generalizations of
D-type Argyres-Douglas theories, which have $SU(2)$ global symmetry, to
theories with arbitrary $A,D$ and $E$ global symmetries).
Also the ${\cal N}=2$ vacuum geometry for some theories we study is captured by an LG period geometry rather than a curve or a local Calabi-Yau 3-fold.   The appearance of mirror geometries which are not Calabi-Yau is familiar from the mirror symmetry story \cite{Vafa:1991uz}.

On the other hand, for the same class of theories we can tune the parameters so that $\tau$ is no longer an exact marginal 
deformation of the 4d theory.  Surprisingly, we find using mirror symmetry an emerging punctured Riemann surface over which there is an $ADE$ type singularity.  This curve is
nothing but the punctured Riemann surface of  class ${\cal S}$ construction \cite{Gaiotto:2009we,Gaiotto:2009hg}. Using our mirror geometry, we identify the puncture type
for a large class of examples.  We also verify the conjecture 
presented in \cite{Ohmori:2015pua} for a number of highly non-trivial cases. For this limit of 6d compactification, the S duality group is interpreted as the mapping class group of this emerging punctured Riemann surface.

The lesson we learn from these two roads is that totally different 4d theories could have a single 6d origin. The compactification leads to different 
theories depending on what kind of property we want to keep in lower dimension. For the first class of theories, the flavor symmetry is broken and
it shows up in the moduli space of the 4d theory, and the conformal theory skips dimension 5.  In the second class, the global symmetries of the 6d are preserved but
the geometry of $T^2$ and its $SL(2,\Z)$ symmetry is irrelevant, and there is a 5d CFT parent.

We also study other examples including toroidal compactification of $A$-type 6d conformal matter.
In M-theory, this corresponds to $M$ M5 branes probing an $A_{N-1}$ singularity, and in the tensor branch it
is a linear quiver with gauge group $SU(N)^{M-1}$.  We show that by compactifying this theory on $T^2$ and tuning
parameters appropriately we can get an arbitrary punctured genus $g$ theory of class $\cS[A_k]$ (where $g$,$k$
depend on $N,M$.)  In this case we land on a restricted class of curves for which the $SL(2,\Z)$ symmetry of the torus
acts as part of the mapping class group. We also show that the little string version of these theories lands us on the class $\cS$ theories
of $A$-type with no punctures.

The organization of this paper is as follows: in section 2 we briefly discuss the compactification of 6d theories which arise from M5 branes probing an $A_{N-1}$ singularity.  This simple example illustrates many of the salient features of the more intricate systems which are the focus of the present paper. In section 3 we review the main character of our play: the 6d SCFTs which in F-theory geometry correspond to orbifold elliptic CY 3-folds and
its compactification to 5 and 4 dimensions.  
Section 4 reviews how to write down 
the Landau-Ginzburg mirror for the toroidal compactifications of the orbifold theories which is then identified with the effective
${\cal N}=2$ geometry of the 6d theory on $T^2$; In section 5, we study many explicit
examples including those where the $SL(2,\Z)$ symmetry of the $T^2$ acts as the duality group. In section 6, a different type of 4d SCFT is found and an emerging 
punctured Riemann surface appears whose mapping class group would be the duality group \cite{Gaiotto:2009we}. In section 7 we present brief concluding thoughts.
Some details are discussed in the appendices.

\section{Toroidal compactification of A-type 6d theories}\label{comicbook}

\begin{figure}
\begin{center}
\small
\centering
\includegraphics[width=0.4\textwidth]{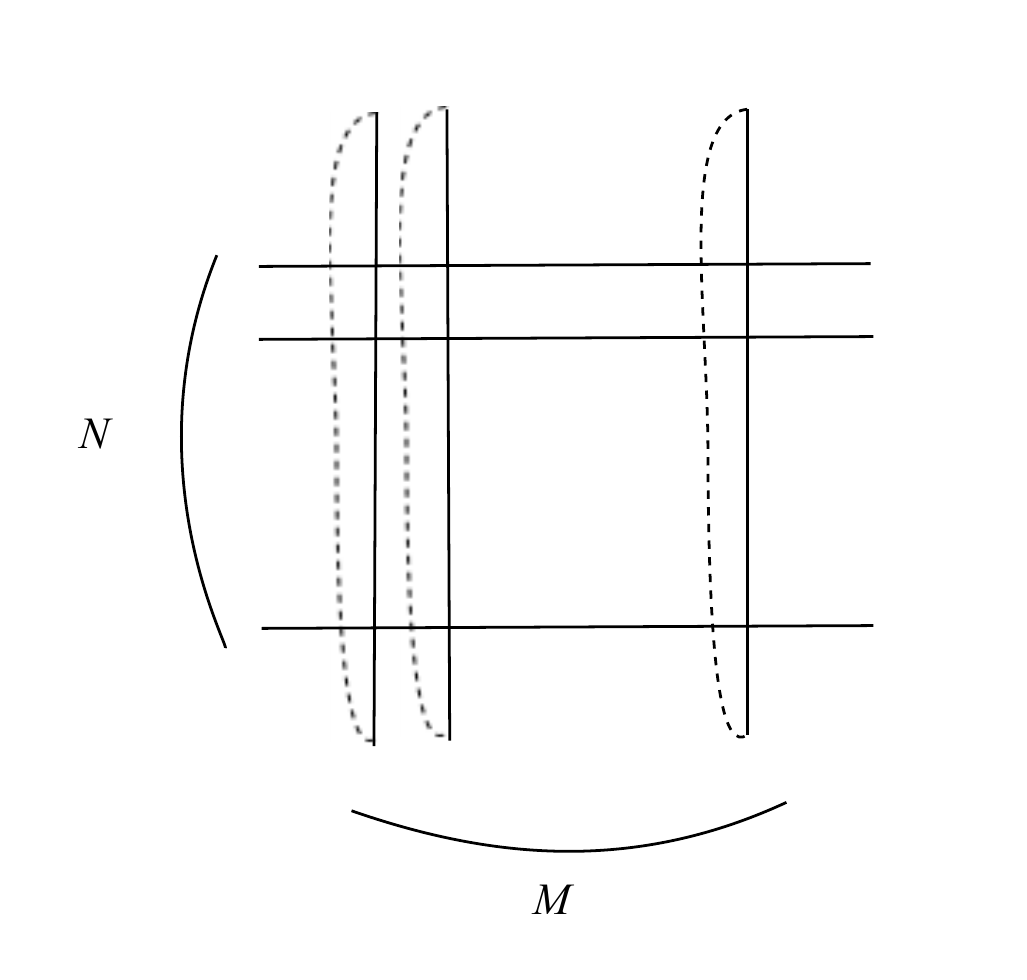}
\caption{Type IIB mirror toric webs for compactification of 6d conformal theories of $A$-type.  Generic situation which upon compactification gives rise to a Seiberg-Witten curve on a genus $g= M + (M-1)(N-1)$ Riemann surface with $2N$ punctures.}
\label{toricweb1}
\end{center}
\end{figure}

\begin{figure}
\begin{center}
\small
\centering
\includegraphics[width=0.4\textwidth]{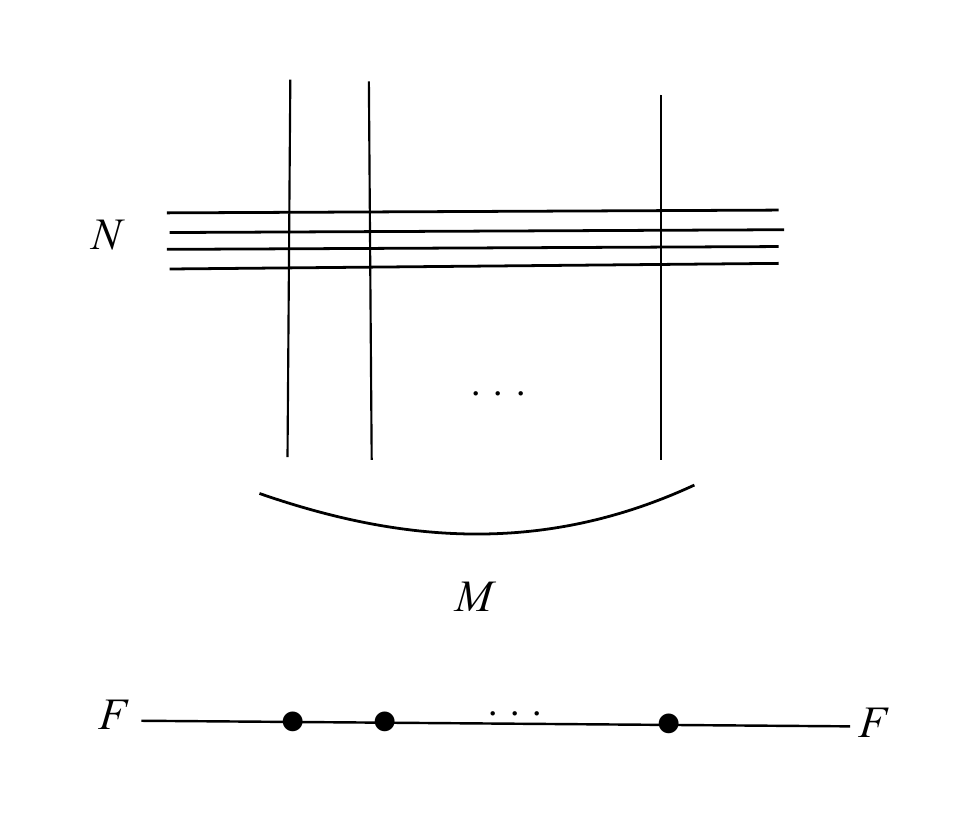}
\caption{\textsc{up:} brane web limit (no torus) \cite{Hanany:1997gh,Brunner:1997gf} \textsc{down:} corresponding degeneration limit of a sphere with $M$ simple punctures and 2 full punctures. }
\label{toricwebnotorus}
\end{center}
\end{figure}

In this section we briefly discuss some aspects of compactification of the SCFT that in M-theory arises
by considering M M5 branes probing an $A_{N-1}$ singularity. Upon compactification on $S^1$ a dual description of this theory can be given \cite{Haghighat:2013gba,Hohenegger:2013ala,Haghighat:2013tka} in terms of M-theory on certain Calabi-Yau manifolds  or equivalently
$(p,q)$ web of 5-branes in IIB theory on a 2d plane where one direction of the plane is compactified on a circle.
We get a toric geometry which looks as in figure \ref{toricweb1}.   Such toric geometries were considered originally in \cite{Hollowood:2003cv}.
It was shown there that as we go on down on another circle, where we obtain the dual type IIA setup, the mirror type IIB
geometry is given by
$$\sum_{r=0}^M\prod_{i=1}^N a_r \vartheta(x-u^r_i,\tau)\ y^r = uv$$
or equivalently the Seiberg-Witten curve is
$$\sum_{r=0}^M\prod_{i=1}^Na_r \vartheta(x-u^r_i,\tau)\
y^r=0$$
where $y=exp(-Y)$ is a $\C^*$ variable, $x$ takes its values on the torus given by a complex
parameter $\tau$ and $\vartheta$ denotes the usual Jacobi theta function (where $\vartheta(0,\tau)=0$).  Moreover,
there is a restriction $\sum_i u^r_i=u$ is independent of $r$.   The $\tau$ appearing here is the same as the complex
structure of the $T^2$ which compactifies the 6d theory down to 4d. 
The question is which 4d theories does this lead to.  The most obvious limit to take, by turning off the $u^r_i$ and expanding
the theory near $x=0$ gives the SW curve
$$\sum_{r=0}^M a_r x^r y^r =0$$
which is the conformal point associated to the linear quiver of $SU(N)^{M-1}$ with extra fundamental matters at the two ends. 
This is as expected the most naive reduction of the 6d theory which itself can be viewed, in the tensor branch, as such a quiver theory  (see figure \ref{toricwebnotorus}).  Note that this reduction preserves the $SU(N)\times SU(N)$ flavor symmetry of the 6d theory. This setup should generalize to all models of \cite{Gaiotto:2014lca} which correspond to adding Nahm pole boundary conditions in a massive type IIA setup or, equivalently, $T$-branes in an $F$ theory engineering \cite{DelZotto:2014hpa}. These models in 6d correspond to a linear quiver with decorations on the sides characterized in terms of embeddings of $\mu_L,\mu_R \colon \mathfrak{su}_2 \to G$, encoding the flavor symmetry. It is obvious that the two full punctures in figure \ref{toricwebnotorus} gets replaced by two punctures labeled by $\mu_L$ and $\mu_R$ respectively.

\begin{figure}
\begin{center}
\small
\centering
\includegraphics[width=0.5\textwidth]{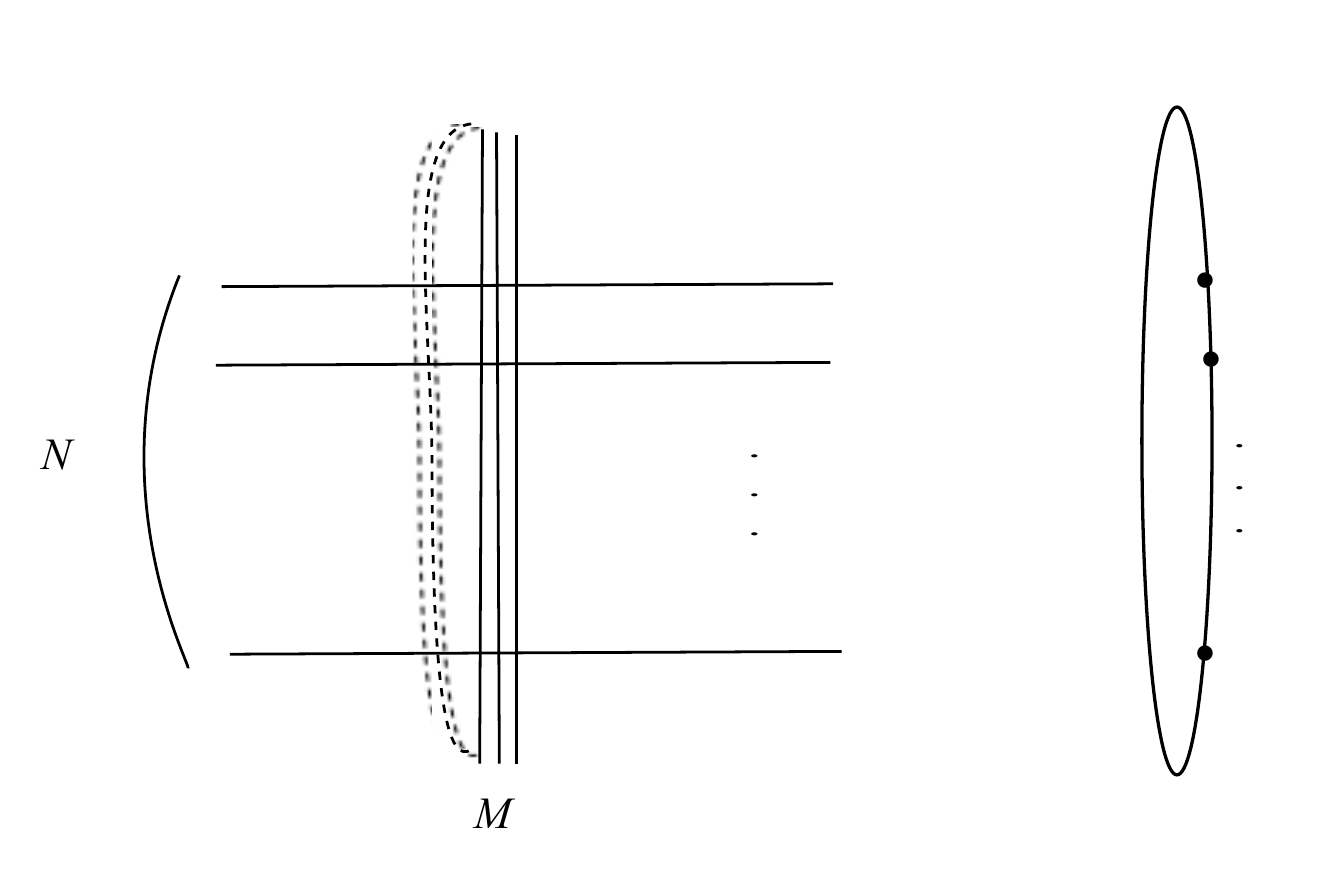}
\caption{ Horizontal/vertical (fiber/base) duality and corresponding degeneration limit: class $\cS$ $A_{M-1}$ theory on a torus with $N$ simple punctures. }
\label{ellipticA}
\end{center}
\end{figure}

On the other hand there are more interesting reductions one can consider. The first interesting remark is that by viewing the vertical lines as D5 branes and the horizontal lines as the NS5 branes (which is the S-dual interperation from the configuration in figure \ref{toricweb1}), we can obtain the elliptic models of \cite{Witten:1997sc}, which are given by an affine $\widehat{A}_{N-1}$ quiver with $SU(M)^{\otimes N}$ gauge group (see figure \ref{ellipticA}). Note that the moduli space of these theories, as pointed out in \cite{Witten:1997sc} is the same as the moduli space of
$N$ points on $T^2$.  This can also be viewed as moduli space of $SU(N)$ flat connections on $T^2$, which in this form finds
a natural interpretation in $6d$:  The $SU(N)$ flat connection is the Wilson line associated with the diagonal $SU(N)_D\subset
SU(N)\times SU(N)$ flavor symmetry of the 6d theory, which one can turn on over $T^2$.  In particular the $6d$ flavor
symmetry is completely broken in this limit.

Already with this first example, we see that we can get two very different 4d $\cN=2$ theories by considering suitable limits of the 6d theory: one with a large flavor symmetry without an $SL(2,\Z)$ symmetry, and the other with a manifest $SL(2,\Z)$ action at the conformal point but with no flavor symmetry. Moreover, in going from figure \ref{toricweb1} to \ref{toricwebnotorus} to \ref{ellipticA} the effective $4d$ theory jumped several times: from a generic SW curve on a genus $g= M + (M-1)(N-1)$ Riemann surface with $2N$ punctures to $\cS[A_{N-1}]$ on a sphere with $M$ simple punctures and 2 full punctures to $\cS[A_{M-1}]$ on a torus with $N$ simple punctures. 

\begin{figure}
\begin{center}
\small
\centering
\includegraphics[width=0.4\textwidth]{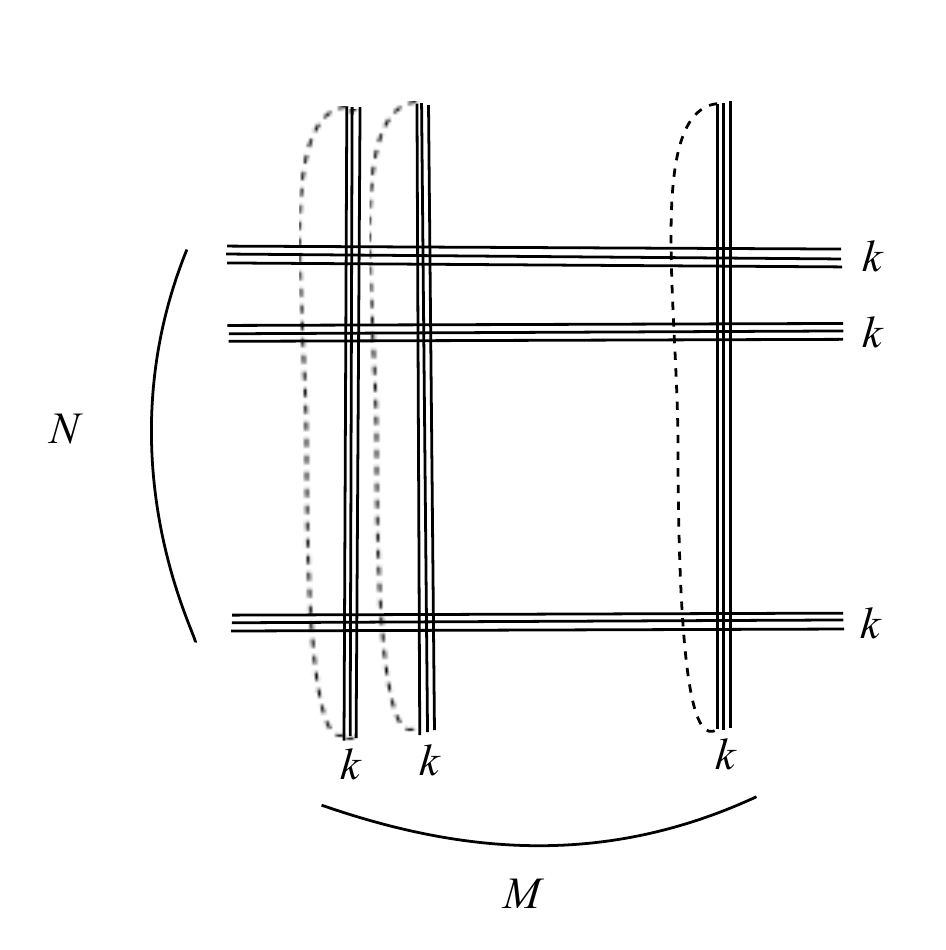}
\caption{Brane web configuration giving the 5d version of a theory of class $\cS[A_{k-1}]$ on a genus $g=g^\prime + (g^\prime-1)(p-1)$ Riemann surface with $2p$ full punctures, where $M=g^\prime k$ and $N=p k$ }
\label{toricweb2}
\end{center}
\end{figure}

In figure \ref{toricweb2} we show that we can obtain in facts all ${\cal N}=2$ theories of class $\cS$ of type $A_{k-1}$  with $2p$ punctures
on a Riemann surface of genus $g=g^\prime+(g^\prime-1)(p-1)$ where $g^\prime$ and $p$ are chosen such that
$$M=g' k,\qquad N=pk.$$

\begin{figure}
\begin{center}
\small
\centering
\includegraphics[width=0.9\textwidth]{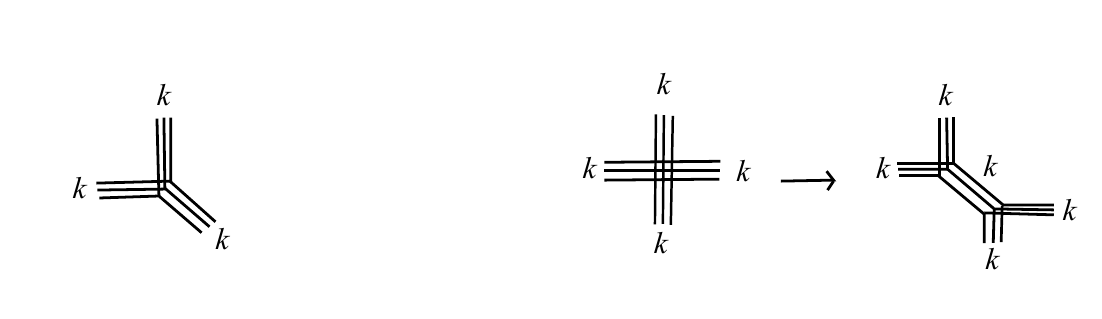}
\caption{Examples of 5d versions of class $\cS[A_{k-1}]$ theories. \textsc{left:} 5d $T_k$ theory; \textsc{right:} 5d lift of class $\cS[A_{k-1}]$ on a sphere with 4 full punctures and corresponding realization of it as the glueing of two $T_k$ theories.}
\label{5dTN}
\end{center}
\end{figure}

This can be anticipated by recalling the 5d lift of class $\cS[A_{k-1}]$ theory \cite{Benini:2009gi, Kozcaz:2010af, Kozcaz:2010yp, Hayashi:2013qwa, Bao:2013pwa , Aganagic:2013tta,Aganagic:2014oia,Aganagic:2014kja,Kim:2014nqa} (see figure \ref{5dTN}). Indeed, as noted in \cite{Vafa:2012fi}, different class $\cS$ theories in 4d can be obtained from the same 5d CFT.  In our case
we group the horizontal lines to $p$ groups of $k$ lines and we group the periodic vertical
lines to $g'$ groups of $k$ lines.  It is not too difficult to see from the geometry that we get a genus $g$ curve which is a $g'$-fold cover
of the $T^2$ together with $2p$ punctures.  This can also be seen from the SW curve as the locus of the curve given by
$$f(x,y)^k=0$$
where
$$f(x,y)=\sum_{r=0}^{g'} a_r \prod_{i=1}^p \vartheta(x-u_i^r,\tau) y^r$$
One can recognize $f(x,y)=0$ as defining a genus $g$ curve (viewing $y$ geometry as a $g'$-sheeted
cover of $T^2$), together with $2p$ full-punctures corresponding
to $y\rightarrow 0$ and $y\rightarrow \infty$ of the above geometry:
$$y\rightarrow 0: \qquad x=u_i^0\qquad i=1,...,p$$
$$y\rightarrow \infty: \qquad x=u_i^{g'} \qquad i=1,...,p$$

\begin{figure}
\begin{center}
\small
\centering
\includegraphics[width=0.4\textwidth]{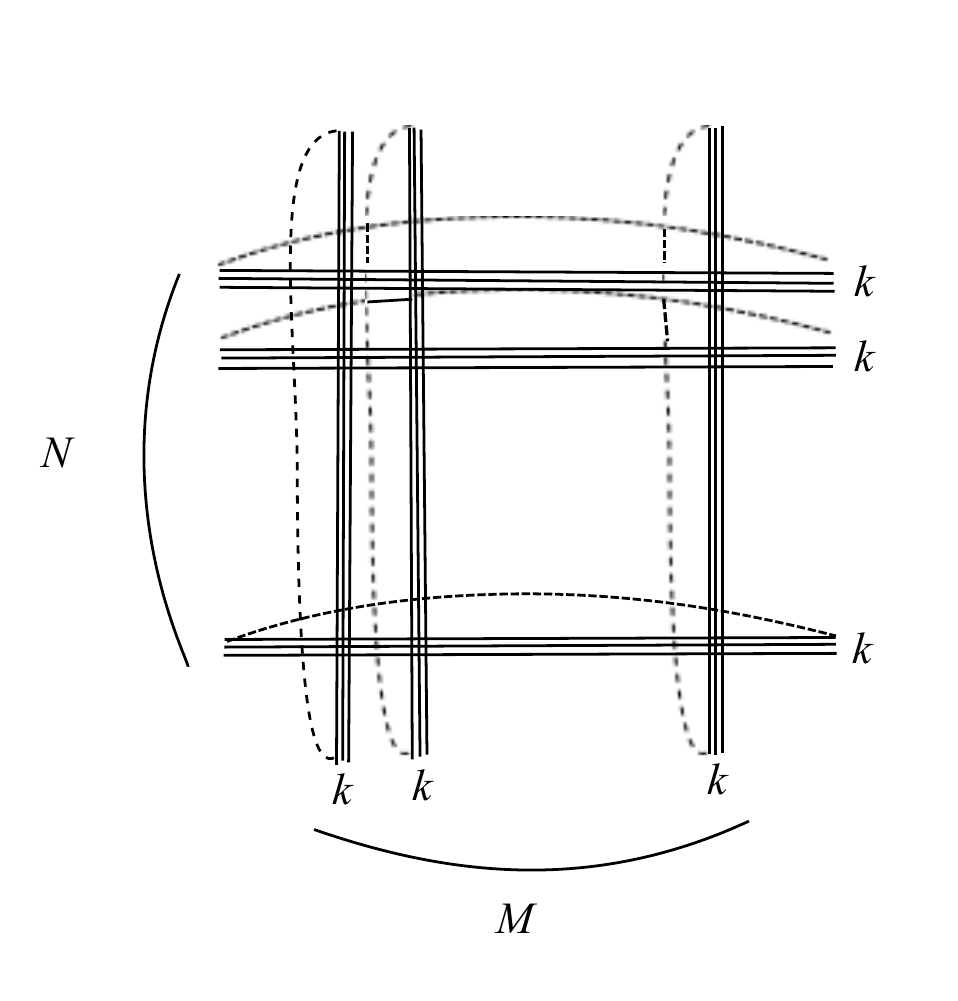}
\caption{5d version of a theory of class $\cS[A_{k-1}]$ on a genus $g=p+g^\prime + (g^\prime-1)(p-1)$ Riemann surface without punctures (toroidal compactification of a little string theory).}
\label{toricweb3}
\end{center}
\end{figure}

Note that $f(x,y)=0$ gives a special type of genus $g$ curve, and not the most
general complex structure.  This is an analog of the `swampland' scenario \cite{Vafa:2005ui} for the field theory setup:  A given
QFT can be consistent in $d$ dimension, but only a subset (or with some restrictions
on their moduli spaces) can arise from $d+k$ dimensional theories
with a given SUSY.  In other words, adding extra degrees of freedom in the UV to a given QFT may or may not
lead to consistent higher dimensional theory.  Thus the purely field theory version of the swampland question
is which field theories do admit such a completion to higher dimensional quantum theories without gravity and with a given amount of supersymmetry.

It is amusing to note that we can also obtain a theory in 4d of $A$-type class ${\cal S}$ with no punctures by considering
the little string theory \cite{Aharony:1998ub,Aharony:1999ks} of the above setup \cite{Haghighat:2013tka} (see also \cite{Kim:2015gha}).  In this case the toric geometry is doubly periodic and we end up periodically identifying horizontal space as well. In the above set up, this
is equivalent to gluing the left and right punctures together and obtaining a theory on a genus
${\tilde g}=pg'+1$ curve with no punctures (see figure \ref{toricweb3}).

\section{6d SCFTs}

The classification of 6d SCFTs \cite{Heckman:2013pva,Heckman:2015bfa} is based on their geometric engineering in F-theory. The corresponding F-theory geometry giving rise to a 6d SCFT involves elliptic CY 3-folds with local singularities.
Some of the singularities may be manifest in the 2 complex dimensional base $B$ of the 3-fold.  Others are hidden in the information of how the elliptic fiber completes the geometry of the 3-fold.  The singularity types of the base were classified in \cite{Heckman:2013pva} and it was found that they are all orbifold singularities embedded in $U(2)$, generalizing the ADE case which embeds in $SU(2)$ and leads  the $(2,0)$ theory as a subclass of the $(1,0)$ SCFTs.
However, the full elliptic Calabi-Yau threefold is not in general an orbifold, because
the elliptic fibration is not in general as simple as would be the case for orbifolds.  Nevertheless
a large class of examples exist which are full elliptic 3-fold orbifolds of $T^2\times \C^2$,  which will be a major focus for the rest of this work.

\subsection{Orbifold 6d SCFTs}  
Let $X=T^2\times {\C}^2$ and consider an orbifold of it $X/\mathsf{G}$ leading to a CY 3-fold.  Such a $\mathsf{G}$ is a subset $\mathsf{G}\subset U(1)\times SU(2)\subset SU(3)$,
where we view each element of $\mathsf{G}$ as a $3\times 3$ matrix
$$\left(\begin{matrix}\alpha^2 & \\ & \alpha^{-1} g\end{matrix}\right)$$
where $g\in \Gamma$ is an element of a discrete subgroup $\Gamma_{ADE}\subset SU(2)$ and where we restrict $\alpha$ such that $\alpha^2$ is an element of $\Z_k$ where $k=2,3,4,6$ in order to be an isomtery of $T^2$.   The choice of $T^2$ complex moduli is restricted:  For $\Z_2$ there is no restriction, for $\Z_3, \Z_6$ we have the hexagonal torus with $\tau=exp(2\pi i /3)$
and for $\Z_4$ we have a square torus with $\tau =i$.   A simple example is  $\mathsf{G}=\langle \Gamma_{ADE}, \Z_{2k}\rangle $ (up to a $\Z_2$ quotient if the center of $SU(2)$ is in $\Gamma$ this is the same as $\Gamma_{ADE}\times \Z_{2k}$).
F-theory on $X/\mathsf{G}$ gives rise to a $(1,0)$ SCFT
in 6d.  There are two ways that $\mathsf{G}$ can have non-trivial elements with one eigenvalue being 1.
If the $1$ is in the fiber $T^2$ direction, this leads to an element of an ADE subgroup $\Gamma_{ADE}$ discussed
above.  If the eigenvalue 1 is in one of the other two directions then the elements of that form will be of the
type $(a;a^{-1},1) $, where $a^p=1$ with $p=2,3,4,6$.  In such a case the base B of the 3-fold
which is the visible part of the space to IIB, will have a line of singularity. It is already known from \cite{Sen:1996vd,Dasgupta:1996ij}
that such singularities of F-theory lead to gauge symmetries $H=SO(8),E_6,E_7,E_8$ in 8 dimensions respectively.
Therefore having these lines of singularities in the 6d case will lead to global symmetries involving these groups.  More
precisely, if the projection of $\mathsf{G}$ in the $T^2$ direction is $\Z_k$ it will lead to these groups or their quotients by outer automorphisms of order $k/p$, namely $H/\Z_{k/p}$.
In particular we get
\be\label{nonsimplylcd}
\begin{tabular}{c|cccc}
&$k=2$&$k=3$&$k=4$&$k=6$\\
\hline
$p=2$&$SO(8)$&$-$&$SO(7)$&$G_2$\\
$p=3$&& $E_6$&$-$&$F_4$\\
$p=4$&&&$E_7$&$-$\\
$p=6$&&&&$E_8$
\end{tabular}
\ee
We can also get more than one eigenvalue of 1 in the $\C^2 $ directions, in which case we will
get a product group as the flavor symmetry.

The singularity of $X/\mathsf{G}$ can be partially resolved by blow up in the base $B$.  The general structure will involve
a collection of spheres in the form of specific type of trees, together with some gauge
group on some of the ${\PP}^1$'s resulting from the Kodaira fiber singularities of the elliptic fiber.
Below we consider some examples which will be useful for us.  This will lead to a theory with $T$ spheres.
$T$ also counts the number of tensor multiplets whose scalar component controls the size of the corresponding sphere.
Also we have gauge group $\prod_{i=1}^T G_i$ (where some $G_i$ may be trivial if they are on spheres
with negative self-intersection 1,2) and some flavor group $G_F=\prod_{i=1}^f H_i$. Notice that the size of the corresponding sphere controls the corresponding gauge coupling as well.
For future notation we denote by
$$r_G=\sum_{i=1}^T rank(G_i)$$  
$$r_F=\sum_{i=1}^f rank(H_i)$$
In what follows we describe the orbifold models in more detail, as well as their $F$--theory geometry. We adopt the notation of \cite{DelZotto:2014hpa,Heckman:2015bfa}, where the structure of the tensorial Coulomb branch of a given 6d SCFT is represented as follows:
$$
[F] , \overset{\mathfrak{g}_1}{n_1}, \overset{\mathfrak{g}_2}{n_2}, \dots 
$$
where the notation $[F]$ means that $F$ is a flavor symmetry, a non--compact divisor supporting a singularity of type $F$, while the notation $ \overset{\mathfrak{g}}{n}$ stands for a compact $\PP^1$ with self intersection $-n$ supporting a singularity of type $\mathfrak{g}$.  Wrapping $D3$ branes on such $\PP^1$ gives rise to a tensionless string in the 6d theory as we shrink the $\PP^1$.  $\mathfrak{g}$ encodes the type of $7$-brane giving rise to the 6d gauge group. The matter content can be determined from this datum using 6d gauge anomaly cancellation \cite{Green:1984bx,Erler:1993zy,Sadov:1996zm,Bershadsky:1997sb,Blum:1997fw,Riccioni:1998th,Grassi:2011hq,Heckman:2015bfa}.

\subsubsection{$\cO(-n)$ models, i.e. $\mathsf{G}=\{ (\alpha^2;\alpha^{-1},\alpha^{-1})\}$ and $\alpha\in \Z_{n=2,3,4,6,8,12}$}\label{originalorbfs}
These cases were originaly studied in \cite{Witten:1996qb} and they correspond, after the blow up
which gets rid of the singularity, to a single $\PP^1$ with negative self-intersections 2,3,4,6,8 and 12, respectively. Moreover the elliptic fibration, except for the $\Z_2$ case which leads to the $A_1$ (2,0) theory,
has some singularity leading to gauge symmetry on them. 
\begin{table}
\begin{center}
\begin{tabular}{cccccccc}
\phantom{$\Bigg|$  } $\cO(-2)$ & $\cO(-3)$&$\cO(-4)$&$\cO(-6)$&$\cO(-8)$&$\cO(-12)$ \\
\hline
\phantom{$\Bigg|$}$2$& $\overset{\mathfrak{su}_3}{3} $&$ \overset{\mathfrak{so}_8}{4} $&$ \overset{\mathfrak{e}_6}{6}$&$\overset{\mathfrak{e}_7}{8} $& $\overset{\mathfrak{e}_8}{12}$\\
\end{tabular}
\caption{Structure of the tensorial Coulomb branches of the $\cO(-n)$ models.}
\end{center}
\end{table}
Here we will get the generic singularities which lead
to gauge groups $SU(3),SO(8), E_6,E_7$ and $E_8$ for the cases $3,4,6,8,12$ respectively. These models are very interesting as these are among the simplest 6D (1,0) tensor--vector systems which have a single tensor and are non-Higgsable as well \cite{Bershadsky:1997sb,Morrison:2012np}. Many examples that we consider here can be considered as orbifolds of these theories (see section \ref{Oncousins}).

\subsubsection{$\mathsf{G}=\langle \Z_k, \Z_{Nk}\rangle,\ k=2,3,4,6$ and $(G,G)$ conformal matter}\label{orbfsandconfmat}
Let us next consider the case in which $\mathsf{G}$ is generated by two elements
\be\label{conformattorbofo}
g=(a;a^{-1},1), \ h=(1;b,b^{-1})
\ee
where $a,b$ are primitive roots of unity with $a^k=1$ and $b^{Nk}=1$.  Note that $g h^N$ will lead
to the element $(a; 1,a^{-1})$ and so this theory will enjoy the symmetry $D_4\times D_4, E_6\times E_6,E_7\times E_7,
E_8\times E_8$ for $k=2,3,4,6$ respectively. Indeed this theory corresponds to $\cT(G,N)$, the theory of N conformal matter of $D_4,E_6,E_7,E_8$ type along a linear chain \cite{DelZotto:2014hpa,Heckman:2014qba}, which also arises in M-theory from N M5 branes probing $D_4,E_6,E_7 $ and $E_8$ singularities. To show that this is indeed the case, note that for $N=1$ the base $B$ is not singular because it can be viewed as $\C/Z_k \times \C/Z_k$ which by
a change of coordinates is isomorphic to $\C\times \C$.  Modding by $\mathsf{G}$ corresponds to modding this geometry by an additional $\Z_N$
which leads to an $A_{N-1}$ singularity of type IIB, which makes contact with the M-theory description involving $N$ M5 branes. So we have two non-compact divisors supporting $G$ flavor symmetry, colliding at an $A_{N-1}$ singularity, which is precisely the setup of  \cite{DelZotto:2014hpa} (see also \cite{Bershadsky:1996nu,Aspinwall:1997ye}).

Let us proceed by reviewing the structure of the corresponding tensorial Coulomb branches of these types of conformal matter.
\begin{itemize}
\item The tensorial Coulomb branch of the $\cT(SO(8),N)$ theory is:
\be
[SO(8)],1,\overset{\mathfrak{so}_8}{4},1,\overset{\mathfrak{so}_8}{4},1,\overset{\mathfrak{so}_8}{4},\dots,1,[SO(8)]
\ee
which has $T = 2N-1$, $r_G = 4(N-1) $ and $r_F = 8$;
\item The tensorial Coulomb branch of the $\cT(E_6,N)$ theory is:
\be
[E_6],1,\overset{\mathfrak{su}_3}{3},1,\overset{\mathfrak{e}_6}{6},1,\overset{\mathfrak{su}_3}{3},1,\overset{\mathfrak{e}_6}{6},\dots,1,\overset{\mathfrak{su}_3}{3},1, [E_6]
\ee
the resulting theory has $T = 4N - 1$, $r_G =8N-6$, and $r_F = 12$;
\item  The tensorial Coulomb branch of the $\cT(E_7,N)$ theory is:
\be
[E_7],1,\overset{\mathfrak{su}_2}{2},\overset{\mathfrak{so}_7}{3},\overset{\mathfrak{su}_2}{2},1,\overset{\mathfrak{e}_7}{8},1,\overset{\mathfrak{su}_2}{2},\overset{\mathfrak{so}_7}{3},\overset{\mathfrak{su}_2}{2},1,\overset{\mathfrak{e}_7}{8},\dots,1,[E_7]
\ee
with $T = 6N-1$, $r_G = 12N-7$, and $r_F = 14$;
\item The tensorial Coulomb branch of the $\cT(E_8,N)$ theory is obtained by glueing together $N$ copies of the $(E_8,E_8)$ conformal matter 
\be
[E_8]1,2,\overset{\mathfrak{sp}_1}{2},\overset{\mathfrak{g}_2}{3},1,\overset{\mathfrak{f}_4}{5},1,\overset{\mathfrak{g}_2}{3},\overset{\mathfrak{sp}_1}{2},2,1,[E_8]
\ee
along a linear chaing by gauging the adjacent $E_8$'s. Therefore we get $T=12N-1$, $r_G = 18N-8$ and $r_F = 16$.
\end{itemize}

\subsubsection{$\mathsf{G}=\{g^m| g=(\alpha^{-q-1};\alpha, \alpha^{q})\}$ and 6d non-Higgsable theories}\label{subsaborbfs}
This is the case where $\mathsf{G}$ is an order $p$ cyclic group
where $\alpha=exp(2\pi i/p)$ and $p$ and $q$ are relatively prime.  Of course we need 
\be\label{thecondition}
k(q+1)=0 \text{ mod } p,
\ee
for this to respect an isometry of a $T^2$, where $k=2,3,4,6$.  The blow up geometry
of this class of examples has been worked out \cite{Heckman:2013pva} (see also appendix B of \cite{DelZotto:2014hpa}). In particular when one blows down all the spheres
with self-intersection -1 one gets a chain of spheres with negative self-intersections $n_1,n_2,n_3,...,n_r$, where
$${p\over q}=n_1-{1\over n_2-{1\over n_3 -...{1\over n_r}}}$$
In table \ref{genaborb} we list all possible bases which are compatible with this condition. The structure of the tensorial Coulomb branches of these systems is easily obtained from the algorithm above, as these are non-Higgsable of the type classified in \cite{Heckman:2013pva}. 

\begin{table}
\begin{center}
\begin{tabular}{cccc}
endpoint & $p$ & $q$ & $k$\\
\hline
\phantom{$\Big|$}$7 ,A_N ,7$ & $36 N + 48$ &$6N+7$ & $6$\\
\phantom{$\Big|$}$2,2,2,2,3,A_N,3,2,2,2,2$ & $36N + 96$ & $30N + 79$ & $6$\\
\phantom{$\Big|$}$7,A_N,3,2,2,2,2$ & $36N + 72$ & $6N + 11$ & $6$\\
\phantom{$\Big|$}$2,2,2,2,3,A_N,7$ & $36N + 72$ & $30N + 59$ & $6$\\
\phantom{$\Big|$}$5, A_N, 5 $& $16 N + 24$ &$4N+5$ & $4$ \\
\phantom{$\Big|$}$2,2,3,A_N,3,2,2$ & $16N + 40$ & $12N+29 $ & $4$\\
\phantom{$\Big|$}$2,2,3,A_N,5$ & $16N + 32$ & $12N+ 23$ & $4$\\
\phantom{$\Big|$}$5,A_N,3,2,2$ & $16N + 32$ & $4N+7 $ & $4$\\
\phantom{$\Big|$}$4, A_N ,4$ & $9 N + 15$ & $3N+4$ & $3$\\
\phantom{$\Big|$}$2,3 ,A_N, 3,2$ & $9N+21$ & $6N+13$& $3$\\
\phantom{$\Big|$}$4 ,A_N, 3,2$ & $9N+18$ & $3N+5$& $3$\\
\phantom{$\Big|$}$2,3 ,A_N, 4$ & $9N+18$ & $6N+11$& $3$\\
\phantom{$\Big|$}$3, A_N, 3$ &$4 N + 8$&$ 2N+3$ & $2$ \\
\hline
\phantom{$\Big|$} $2,2,2,2,4,2,2,2,2$ & 60 & 49& 6\\
\phantom{$\Big|$} $2,2,2,3,2,2,2$ & 24 & 19 & 6\\
\phantom{$\Big|$} $8,2,2,2,2$ & 36 & 5 & 6\\
\phantom{$\Big|$} $2,2,2,2,8$ & 36 & 29 & 6\\
\phantom{$\Big|$} $2,2,4,2,2$ & 24 & 17 & 4\\
\phantom{$\Big|$} $6,2,2$ & $16$ & $3$ & 4\\
\phantom{$\Big|$} $2,2,6$ & $16$ & $11$ & 4\\
\phantom{$\Big|$} $2,3,2$ & $8$ & $5$ & 4\\
\phantom{$\Big|$} $2,4,2$ & $12$ & $7$ & 3\\
\phantom{$\Big|$} $5,2$ & $9$ & $2$ & 3\\
\phantom{$\Big|$} $2,5$ & $9$ & $5$ & 3\\
\hline
\end{tabular}
\end{center}
\caption{Endpoints which are compatible with the condition of eqn.\eqref{thecondition} and corresponding values of $k$.}\label{genaborb}
\end{table}

\subsubsection{Abelian orbifolds of $\cO(-n)$ models}\label{Oncousins}
In this section we consider the orbifold cousins of the $\cO(-n)$ models, we reviewed in section \ref{originalorbfs}. For simplicity we are going to discuss the case in which $\Gamma$ is an abelian subgroup of $SU(2)$. These models have an $F$-theory realization as orbifolds of $T^2 \times \C^2$ where the orbifolding group is generated by two elements:
\be\label{minimorbif}
g = (\omega^{-2} ; \omega,\omega) \quad \omega^n = 1 \qquad h = (1 ; \alpha, \alpha^{-1}) \quad \alpha^r=1
\ee
The easiest case to analyze is the case in which $n$ and $r$ are relatively prime. According to our previous discussion this orbifold action does not have fixed loci in the K\"ahler base of the $F$-theory geometry, therefore these systems are not going to have any flavor symmetry in 6D. Based on this fact, and on the type of singularity which can be obtained from orbifolded tori, we expect that these models are going to be of the non--Higgsable type studied in \cite{Heckman:2013pva}. To show that this is indeed the case we should realize these orbifold groups as $U(2)$ discrete subgroups. As we are focusing on cyclic subgroups, we expect that the models considered in this subsection are all going to be of generalized $A$-type. We proceed by characterizing the corresponding bases. To do that we have to identify the groups generated as in eqn.\eqref{minimorbif} with abelian discrete subgroups of $U(2)$. As $n$ does not divide $r$ by construction, all models of this class are going to have $p = n r$. It remains to determine $q$ in order to be able to reconstruct the corresponding bases from the continued fraction $p/q$. In order to do that we have to solve for an $nr$ root of unity $\xi$ which is such that $\xi = \omega \alpha$ and $\xi^q = \omega \alpha^{-1},$ or, equivalently, to find the least integer $q$ such that $q (n+r) = (n-r) \text{ mod } n r$. Implementing this search systematically we produced the results in table \ref{tablecoprimeorbfs}: obviously, all these models belong to the class we discussed above.

\begin{table}
\begin{center}
\begin{tabular}{ccc}
$n$ & $r$ & endpoint \\
\hline
3 \phantom{$\Big|$}& 4 & 2,4,2 \\
& \phantom{$\Big|$} $3N+1$ & $2,3,A_N,3,2$ \\ 
& \phantom{$\Big|$}2 & 6\\
\phantom{$\Big|$}&$3N +2$ & $4 A_{N-1}4$\\
\hline
4 \phantom{$\Big|$}& \phantom{$\big|$} $4N+1$ & $3 A_{4N-1} 3$\\
 \phantom{$\Big|$}& $4N+3$ & $3 A_{4N+1} 3$\\
\hline
6 \phantom{$\Big|$} & $6N+1$ & $4 A_{4N-1} 4$\\
 \phantom{$\Big|$}& $6N+5$ & $2,3,A_{4N+1},3,2$\\
\hline
8 \phantom{$\Big|$}& $8N+1$ & $5 A_{4N-1} 5$\\
 &\phantom{$\Big|$} $8N+5$ & $5 A_{4N+1} 5$\\
 &\phantom{$\Big|$} 3 & 2,2,4,2,2\\
 &\phantom{$\Big|$} $8N+3$ & $2,2,3,A_{4N-1},3,2,2$\\
 \phantom{$\Big|$}& $8N+7$ & $2,2,3,A_{4N+1},3,2,2$\\
\hline
12 \phantom{$\Big|$}& $12N+1$  & $7 A_{4N-1} 7 $\\
 &\phantom{$\Big|$} 5 & 2,2,2,2,4,2,2,2,2\\
 &\phantom{$\Big|$} $12N+5$  & $2,2,2,2,3,A_{4N-1},3,2,2,2,2$\\
 & \phantom{$\Big|$}$12N+7$  & $7 A_{4N+1} 7 $\\
 &\phantom{$\Big|$} $12N+11$ & $2,2,2,2,3,A_{4N+1},3,2,2,2,2$\\
\end{tabular}
\caption{ Endpoints for coprime orbifolds of minimal models of type $n=3,4,6,8,12$} \label{tablecoprimeorbfs}
\end{center}
\end{table}

Let us proceed by considering the case in which $n$ and $r$ are not coprime. As we shall see, conformal matter are 6D orbifolds of minimal models of this type. Indeed, it is sufficient to choose $r = N n$ and for $n=2,3,4,6$ and these discrete groups and the ones we constructed in section \ref{orbfsandconfmat} are isomorphic. Now, what happens in the cases $n = 8,12$? We claim that one gets back again conformal matters. In the first case one obtains just a collision of two $\Z_4$ singularities corresponding to the fixed loci for the subgroups generated by $g h^N$ and $g h^{-N}$ respectively, which overlap at the origin. At the origin the element $g^4 h$ generates an apparent point of $\Z_{8N}$ singularity, but changing coordinates one has a residual $\Z_{2N}$ singularity at the origin, and therefore we end up with an engineering of the model $\cT(E_7,2N)$. In the case of $n=12$, by the same method one obtains the models $\cT(E_8,2N)$. There are several cases left to analyze. To determine the structure which one obtains in these cases though, requires a direct inspection of the structure of the orbifold groups, which is rather intricate. Let us consider a simple example to illustrate this point: take $n=6$ and $r=3$ above, then we have two loci of $\Z_2$ singularity corresponding to the elements $g^2 h$ and $g^4 h$ which meet at a singularity of type $\Z_6$ generated by $g h^3$, changing coordinates one has a residual $\Z_3$ singularity at the origin, and therefore a model of type $\cT(SO(8),3)$.  We leave the full classification of all possibilities to the interested reader and turn to some instructive examples in the next section.

\subsubsection{$\mathsf{G}=\langle \Z_{k_1},\Z_{k_2} \rangle,\ k_i=2,3,4,6$ and $(G,G^\prime)$ conformal matter}\label{GGprimeconfmatt}
Another class of interesting examples comes from the case of $(G,G^\prime)$ bifundamental conformal matter \cite{DelZotto:2014hpa}. Several such models can be realized as well using the orbifold technique of the present paper. Indeed, one can take $\mathsf{G}$ to be generated by
\be
g = (\alpha,\alpha^{-1},1) \quad \alpha\in \Z_{k_1}\qquad h = (\omega,1,\omega^{-1})\quad \omega\in \Z_{k_2}
\ee
where $k_1$ and $k_2$ take any pairs of values out of $2,3,4,6$ which are mutually compatible as automorphisms of a torus with a given complex structure: recall that $\Z_2$ is compatible with any $T^2$, while $\Z_4$ is not compatible with $\Z_3$ nor $\Z_6$. This leaves us with the following possibilities: $\Z_2$ with $\Z_{2,3,4,6}$, $\Z_3$ with $\Z_{3,6}$, $\Z_4$ with $\Z_4$ and $\Z_6$ with $\Z_6$.   Note that all of these examples, by a change
of variables $(z_1,z_2)\rightarrow (z_1 ^{k_1},z_2^{k_2})$, lead to the base $B=\C^2$, and so they are all very Higgsable models (in the sense of \cite{Ohmori:2015pua}). The case of $\langle \Z_k, \Z_k \rangle$ gives the minimal conformal matter of type $(G,G)$ we have discussed above.   Here we focus on the remaining examples. The two subgroups generated by $g$ and $h$ each correspond to a non-compact divisor supporting a singularity of type $SO(8),E_6,E_7$ and $E_8$. However, in this case, according to our choices of $k_1$ and $k_2$ we will obtain systems of $(G,G^\prime)$ conformal matter (of minimal type) with non-simply laced flavor symmetries. This happens because, the four $\Z_2$ fixed points corresponding to the the factors of the $SU(2)^4$ maximal subalgebra of $SO(8)$ get exchanged by a $\Z_4$ or a $\Z_3$ action, the former giving rise to the $\Z_2$ outer automorphism reducing $SO(8)$ to $SO(7)$, the latter giving rise to the triality automorphism which leaves us with $G_2$ after modding out. Similarly two of the three fixed points of the $\Z_3$ torus which corresponds to the factors of the $SU(3)^3$ maximal subgroup of $E_6$ gets exchanged under a $\Z_6$ action which give rise to $F_4$. This is a concrete realization of the general  discussion around eqn.\eqref{nonsimplylcd} at the beginning of this section. Therefore we obtain the models:
\be
\begin{aligned}
&\mathsf{G} = \langle \Z_2,\Z_3 \rangle \colon (G_2,F_4)\\
&\mathsf{G} = \langle \Z_2,\Z_4 \rangle \colon  (SO(7),E_7)\\
&\mathsf{G} = \langle \Z_2,\Z_6 \rangle \colon (G_2,E_8)\\
&\mathsf{G} = \langle \Z_3,\Z_6 \rangle \colon (F_4,E_8)\\
\end{aligned}
\ee
We recognize here the $(SO(7),E_7)$, $(F_4,E_8)$, and $(G_2,E_8)$ models noted to be very Higgsable in \cite{Ohmori:2015pua}. Let us proceed by showing explicitly that these geometries have the desired features. For $(SO(7),E_7)$ we have indeed that the two fixed loci above meet at the apparent $\Z_2$ singularity generated by $ g h^2 = (1;-1,-1)$.  Then its structure its forced on us by the requirement of very-Higgsability together with our knowledge of its flavor symmetry. 
\be
[E_7],1,\overset{\mathfrak{su}_2}{2},[SO(7)]
\ee
which has  $r_F = 10$, $r_G = 1$, and $T = 2$. Let us proceed with the other models. For the $(E_8,G_2)$ system we obtain that the two flavor divisors meet again at an apparent $\Z_2$ singularity generated by the element $gh^3 = (1,-1,-1)$. The model is
\be
[E_8],1,2,\overset{\mathfrak{sp}_1}{2},[G_2]
\ee
that has $r_F = 10$, $r_G = 1$, and  $T = 3$. Similarly, in the $(E_8,F_4)$ case we obtain an apparent $\Z_3$ generated by $gh^4=(1,\alpha,\alpha^{-1})$ with $\alpha$ a third root of unity.  By very-Higgsability and minimality then the model is
\be
[E_8],1,2,\overset{\mathfrak{sp}_1}{2},\overset{\mathfrak{g}_2}{3},1,[F_4]
\ee
with $r_F = 12$, $r_G = 3$, and $T = 5$.

Our last example is a conformal matter of type $(G_2,F_4)$, we claim that this is the theory of one heterotic $E_8$ instanton \cite{WittenSmall,Ganor:1996mu} in a realization where only $G_2 \times F_4\subset E_8$ is manifest.  More precisely, as we go down on a circle,
this theory becomes dual to the $O(-1)$ theory where we have turned on a Wilson line in the flavor $E_8$ group which
breaks it to  $G_2 \times F_4$.

\subsubsection{More general examples}\label{orbfmoregenex}

Just to see how much more flexibility orbifold construction has, let us consider
another set of examples.  Let us start with an orbifold group generated by
\be
g=(\alpha^{-4},\alpha^3,\alpha) \qquad \alpha \in \Z_{12}.
\ee
In this case, indeed, there is a unique fixed locus which supports a singular fiber. The fixed locus corresponds to the element $g^4 = \text{diag}(\alpha^{-4},1,\alpha^4)$, which gives rise to a $\Z_3$ subgroup with $E_6$ global symmetry. Notice that in addition we have the element $g^3 = \text{diag}(1,i,-i)$ which gives a $\Z_4$ singularity at the origin. In this case, as in the other realization of conformal matter as orbifold singularities, such singularity is only apparent and one has to get rid of it by a suitable change of variables which is dictated by the structure of the flavor divisor to be 
\be
\begin{cases}z_1 \to z_1\\ z_2 \to z_2^3
\end{cases}
\ee
now the element $g^3$ in the new set of coordinates (notice that the coordinate on the $T^2$ fiber has to change as well to compensate) reads $(-1,i,i)$ which correspond to a Hirzebruch-Jung singularity with endpoint $4$. To determine which theory we land on we have to resolve it by the following sequence of blow-ups
\be
[E_6] ,4 \to [E_6], 1, 5 \to [E_6] ,2,1,6 \to  [E_6] ,1,3,1,6
\ee
Therefore we conclude that this system gives the $$[E_6],1,\overset{\mathfrak{su}_3}{3},1,\overset{\mathfrak{e}_6}{6}$$ theory with an $E_6$ global symmetry, which has $T=4$, $r_G = 8$, $r_F=6$. Combining this with our previous findings, we see that we have obtained the systems $$\overset{\mathfrak{e}_6}{6},1,\overset{\mathfrak{su}_3}{3},1,\overset{\mathfrak{e}_6}{6},$$ $$[E_6] 1,\overset{\mathfrak{su}_3}{3},1 [E_6],$$ $$\overset{\mathfrak{su}_3}{3},$$ $$[E_6],1,\overset{\mathfrak{su}_3}{3},1,\overset{\mathfrak{e}_6}{6}$$ all realized as orbifolds! Let us proceed by showing that also the $$\overset{\mathfrak{su}_3}{3},1,[E_6]$$ theory can be realized as an orbifold. We propose that this correspond to the following orbifold action:
\be
g = (\alpha^2; \alpha^3,\alpha) \qquad \alpha \in \Z_6.
\ee 
Indeed, $g^2 = (\alpha^{-2}; 1, \alpha^2)$ is a line of $E_6$ singularity and in this case $g^3 =(1,-1,-1)$ is an apparent $\Z_2$ singularity: the change of variables is the same as before and it maps $g^3$ back to itself, therefore in this case this singularity is a canonical one, and the endpoint geometry is
\be
[E_6] ,2 \xrightarrow{\ \text{blow up} \ } [E_6] ,1, 3
\ee
which concludes our derivation.  These constructions can be combined with other cyclic elements
which live purely in the $SU(2)$ part to give a large number of variation, which will affect the endpoints of the
constructions we have done.  The point of this section was not to do a systematic exploration, but just to illustrate that we can in principle find orbifold examples which are rather rich.

\subsection{Compactifications to 5d}

If we compactify the 6d theory on a circle $S^1$ of radius $R_6$ down to 5 dimensions, the duality between F-theory and M-theory
gives an elliptic threefold description of the theory, where the elliptic fiber of F-theory has K\"ahler class given by $1/R_6$.
In the context of the orbifold SCFTs this leads to M-theory on $(T^2\times \C^2)/\mathsf{G}$.  Of course this is singular
and we can consider blowing up the singularities etc.  In fact we expect the number of Kahler parameters controlling
this geometry to be 
$$L=r_G+r_F+T+1.$$
 This is because upon compactifying the 6d theory on a circle we 
can turn on Wilson lines for gauge and flavor symmetry groups.  In addition to this
we have the original Kahler classes of the $T$ spheres, and one more from the radius $R_6$ of the circle which
gets mapped to the inverse of the Kahler class of $T^2$.

We can also ask if the 6d theory flows to a conformal theory in 5d.  In fact it does, but to many distinct
possible theories, which is familiar in the context of the worldvolume theory of one Heterotic $E_8$ instanton \cite{Seiberg:1996bd,Morrison:1996xf,Douglas:1996xp}: Indeed, from one 6d theory, one obtains the whole family of $E_{N_f+1}$ 5d SCFTs.  For the orbifold SCFTs one obvious place where they would appear is at the singularity
of the geometry.  Note however this is {\it not} the only place they appear.  To explain this more clearly
let us focus on just one example: the $E_6$ conformal matter.  This is the case where 
$\mathsf{G}$ is generated by two elements
$$(\omega;\omega^{-1},1),(1;\omega,\omega^{-1})$$
where $\omega^3=1$.  Note that the global symmetry for this M-theory background can be read off
by looking at the $A_2$ singularities.  Let us label the coordinates of the two complex planes by $(z_1,z_2)$.
Also let $p_i$ denote the three fixed points of $T^2$ under the $Z_3$ rotations as $i=1,2,3$.
We find that we have an $A_2$ type singularity along $z_1$
$$(p_i;z_1,0)$$
as well as along $z_2$
$$(p_i;0,z_2)$$
In the 6d case we had an $E_6$ singularity along each of the $z_i$, but
now each one of these have split to three singularities of type $A_2$.
This implies that each of the two $E_6$ global symmetries has broken to
$$E_6\rightarrow SU(3)\times SU(3)\times SU(3)$$
In other words from the 6d perspective we must have turned on a discrete
$\Z_3$ holonomy leading to this breaking.
This was already noted in \cite{Dasgupta:1996ij} as a generalization of orientifold construction to the F-theory setup.
The same  idea works for the $D_4,E_7,E_8$ as well.  For example in the $D_4$ case
we get four fixed points of the $\Z_2$ action which signifies the breaking of $SO(8)$ to $SU(2)^4$.
In fact this is exactly as one would expect in orientifold constructions, except that we seem to have
gone down in the dual M-theory description on two directions, even though we only compactified one circle.

Given this interpretation it seems natural to expect that we can turn off the Wilson line
and restore the bigger symmetry group, similar to the Polchinski-Witten construction of $E_8$ gauge symmetry
in type I' theory \cite{Polchinski:1995df}.  This would correspond in this language to blowing
up the singularity and going to a suitable corner of moduli space.  Even though it is possible in principle
to do this geometrically, it turns out to be easier to see how this comes about when
we compactify further to 4 dimensions and use mirror symmetry to describe the (complexified) blown up
geometry in terms of complex polynomials.  We will do this in detail later in this paper, and will not systematically study the 5d SCFT fixed points that we flow to.   However, aspects of 5d SCFT's will be useful in shedding light on what we flow to in 4d,
which we now explain.  In particular we concentrate on a different conformal fixed point in 5d, namely the limit where $T^2$ gets big
and we zoom in to any of the three fixed points $p_i$.  Near each of them we simply have the geometry of
$$\C^3/\Z_3\times \Z_3$$ 
which is known to be the 5d analog of $T_3$ theory (which flows upon circle compactification to the 4d $T_3$ theory).
The 5d $T_3$ theory enjoys a manifest $SU(3)\times SU(3)\times SU(3)$ symmetry where each $SU(3)$ comes
from the $A_2$ singularity along any of the three planes of $\C^3$.

Another class of examples we will need involves a local geometric singularity
of M-theory of the form 
$$\C^3/\mathsf{G}$$
with  $\mathsf{G}=\langle Z_{2p}, \Gamma \rangle$ the action on $\C^3$ given by
$$(\alpha^2,\alpha^{-1}\Gamma)$$
where $\alpha^{2p}=1$ and $\Gamma$ is a discrete
subgroup of $SU(2)$, as before.  For this discussion we will not assume any restriction on $p$ (of course
the case of most interest for us would be with special values of $p$ which allow $T^2$ isometries, i.e. $p=2,3,4,6$ where we can
lift this up to 6d).
This theory will be a 5d theory with ADE global symmetry
group $G=ADE$ associated to $\Gamma$ as the flavor symmetry group.  The reason for this is that there is a non-compact locus
of $ADE$ singularity (along the first $\C$ direction).  This 5d theory we will denote by
$${\hat D}_p(G)$$
As we will argue in section 4, upon compactification to 4d this theory flows to ${\cal N}=2$ theories $D_p(G)$  discovered
in \cite{Xie:2012hs,Cecotti:2012jx} as generalization of Argyres Douglas theories of D-type (the usual AD theory
of D-type corresponds to $G=SU(2)$).

If we consider instead, M-theory on a partially compact version of the above geometry, i.e. $(T^2\times \C^2)/ \mathsf{G}$ with $\mathsf{G}$ 
as before, but now with the restriction that $p=2,3,4,6$ to allow $\mathsf{G}$ to act on $T^2$, we will get a number of copies
of ${\hat D}_{m_i}(G)$ theories coming from the fixed points of the $\mathsf{G}$ action on the $T^2$, where $m_i$ are the order of stabilizer of the
corresponding fixed point, and $i$ labels the fixed point.  Moreover now the $G$ is gauged, because the locus of $G$ singularity
is $T^2$ which is compact.  In particular we have
\begin{itemize}
\item  For a $\Z_2$ orbifold of $T^2$, we get four identical $\Z_2$  fixed points, and the 5d theory would consists of a gauge sector with group $G$ gauging the diagonal flavor symmetry of four identical matter systems ${\hat D}_2(G)$; 
\item  For a $\Z_3$ orbifold of $T^2$, we get three identical $\Z_3$  fixed points, corresponding to a $G$ gauge sector coupled to three identical matter systems of type ${\hat D}_3(G)$;
\item For a $\Z_4$ orbifold of $T^2$, we get one $\Z_2$ fixed point and two identical $\Z_4$ fixed points, corresponding to a $G$ gauge sector coupled to 
an ${\hat D}_2(G)$ matter system and two identical ${\hat D}_4(G)$ matter systems.
\item For $\Z_6$ orbifold of $T^2$, we get a $\Z_2$ fixed point, a $\Z_3$ fixed point and a $\Z_6$ fixed point, and the 4d theory would consist of a $G$ gauge group weakly gauging the diagonal flavor symmetry of an ${\hat D}_2(G)$ matter system, an ${\hat D}_3(G)$ matter system and an ${\hat D}_6(G)$ matter system.
\end{itemize}

\subsection{Compactification to 4d}\label{mtheoryengine}

We now consider compactifying the theory on one more circle down to 4 dimensions corresponding to compactifying the 6d theory on $T^2$.  The theory will have ${\cal N}=2$ supersymmetry in 4 dimensions.
In principle we can flow to interesting 4d SCFTs. For this we will have to decouple some modes. In particular the area of the $T^2$ should go to zero.

For the examples discussed in the last section, given by $\mathsf{G} = \langle Z_p,\Gamma\rangle$ as discussed above, we now argue that we end up with an interesting 4d system.  Since upon compactification
$\hat D_p(G)\rightarrow D_p(G)$ (as we will show later) we should get a 4d system which gauges the $G$ global symmetry
of the corresponding $D_p(G)$'s.  In
fact the integers $$\{p_i\}=(2,2,2,2), (3,3,3), (2,4,4),(2,3,6)$$ for the gauged $D_p(G)$ systems which we get from the structure of the torus fixed points are precisely those for which
\be
\sum_i \frac{p_i-1}{p_i} = 2
\ee
which is exactly the condition for which the beta function contribution of gauging the diagonal flavor symmetry of a system of 4d $D_{p_i}(G)$ systems vanishes!  The resulting SCFTs were introduced in \cite{Cecotti:2013lda} as generalization of the findings of \cite{Cecotti:2011rv} in the $G=SU(2)$ case. We denote them by $(E_n^{(1,1)},G)$, $n=4,6,7,8$. In the literature about BPS quivers $E_4^{(1,1)} = D_4^{(1,1)}$ (a.k.a. $SU(2)$ $N_f = 4$). So we should expect that the flow of the 6d SCFT orbifolded by $G$ should flow to this 4d theory upon compactification. An account of many of their interesting properties can be found in appendix \ref{ellipticSCFTs}. One crucial fact about these models is precisley that they enjoy an unexpected (from the 4d perspective) $SL(2,\Z)$ duality symmetry. In the case of the elliptic $(E^{(1,1)}_n,A_1)$ SCFTs obtained by gauging the diagonal $SU(2)$ flavor symmetry of systems of AD $D_{p_i}$ systems, such $SL(2,\Z)$ action was realized explicitly at the level of the corresponding BPS spectrum in \cite{Cecotti:2012va}. Later the $SL(2,\Z)$ duality for the model $(E_7^{(1,1)},A_1)$ has been observed also by \cite{Buican:2014hfa} at the level of the corresponding SW curve. Here we are explaining the origin of this symmetry and predict that such $SL(2,\Z)$ action extends to all $(E_n^{(1,1)},G)$ models. 

For more general theories, we need to find other techniques to analyze the interesting SCFTs we flow to.  Indeed we now recall how in the context
of geometric engineering of ${\cal N}=2$ theories in 4d
mirror symmetry helps us \cite{Klemm:1996bj,Katz:1997eq}.  

Consider compactifying the theory further on a circle of radius $R_5$. We get a description in terms of type IIA on the same
three-fold where the Kahler class of the elliptic fiber is $\kappa=iR_5/R_6$.  If we use mirror symmetry on the elliptic Calabi-Yau,
we land back on a type IIB description which gives an exact description of the quantum corrected ${\cal N}=2$ vacuum geometry. 
In our case, since the IIA geometry involves a $T^2$, the mirror geometry
will also enjoy a $T^2$ fibration structure where $\kappa$ plays the role of the complex structure $\tau$ of the torus. It is worth noting that the $\tau$ which will figure in the mirror type IIB geometry is the complex structure of the torus $T^2$ which we compactify the 6d theory on to get down to 4d.
More generally we will obtain an interesting complex geometry for the type IIB setup which can be used to locate interesting SCFTs.

The complex structure of $T^2$ can sometimes be left at the conformal fixed point as a marginal parameter, in which
case an $SL(2,\Z)$ duality group acts on the 4d theory, as in the case of compactifying $(2,0)$ theories and the $(E_n^{(1,1)},G)$ systems just
discussed, or as we shall
find in some examples, as irrelevant deformations of the 4d theory. When the complex structure $\tau$ is marginal in the 4d theory,
it appears that the parent 5d theory is {\it not} conformal, as is for the $(2,0)$ theory, because to get
to the 5d limit, we need to take $R_5\rightarrow \infty$ which would correspond to $\tau \rightarrow \infty$ which 
is at infinite distance in moduli space.  

In addition a general 6d SCFT may have some non-trivial flavor global symmetry $G$.  In compactifying down we have a choice of how much of $G$
we wish to preserve (e.g. by switching on suitable Wilson lines in going from 6 to 5 dimensions).  As we will see different conformal theories in 4d emerge depending
on this.  For the cases where $G$ is broken, sometimes the flat holonomy of the broken group
on the torus shows up as moduli parameter in 4d.  As we shall see, this structure
explains the appearance of such moduli spaces in certain 4d ${\cal N}=2$ theories. 
 An interesting example is the conformal matter
system $\cT(G,N)$ in 6d corresponding to $N$ M5 branes probing $G$-type singularity.  As already discussed
this theory enjoys $G\times G$ global symmetry.  It turns out in going down to 4d we can preserve $G\times G$ symmetry
or break it and these lead to different conformal systems in 4d, as we discussed for the case of $G=A_N$ in section 2.  In the first instance we end up with the theories
of class ${\cal S}$ with $N$ simple punctures and 2 full punctures generalizing the proposal of \cite{Ohmori:2015pua} for the $N=1$ case.
On the other hand as discussed in \cite{DelZotto:2014hpa} in 5d this same theory is equivalent
to an affine  $ADE$ quiver theory (as generalization of fiber-base duality \cite{Katz:1997eq}) and it naturally
leads to the same theory in 4d which is conformal.  Note that, as pointed out in \cite{Katz:1997eq} the moduli space of this theory is flat $ADE$ connections on a torus.  We can now explain this using the $6d$ picture, namely the diagonal
flavor symmetries on $T^2$ explain this moduli space.  In other words $G\times G$ is completely broken and
the Wilson lines of the diagonal global symmetry $G$ on $T^2$ plays the role of marginal deformations while the other
part of the $G$ flavor symmetry become the mass parameters for the affine theory\footnote{We would like to thank B. Haghighat and G. Lockhart
for discussions on this point.}. Note that in 5d the affine quiver does not lead to a conformal theory.  The reason
for this is clear:  The base of the affine $ADE$ quiver forms an inner product which is not negative definite and so
it cannot all be shrunk to zero at finite distance in moduli space (we can of course shrink all except for the affine node).

\section{Mirror Technology for Orbifolds}

As already discussed we would need to construct the mirror for the type IIA geometries
given by $T^2\times \C^2/\mathsf{G}$.  Here we review the work \cite{Vafa:2001ra} which shows how this can be done explicitly when $\mathsf{G}$ is abelian.
We construct the mirror by constructing the mirror for the $T^2$ and $\C^2$ orbifolds separately
and then combining them.

First consider the geometry $\C^n/\Z_p$ where the action of $\Z_p$ on $\C^n$  is given by
$$(\alpha^{r_1},\alpha^{r_2},...,\alpha^{r_n})$$
where $\alpha^p=1$.  The mirror of this geometry including turning on twistor sector chiral fields is given
by a Landau-Ginzburg theory with two $\C^*$ variables $y_i=exp(-Y_i)$ with superpotential
$$W=y_1^p+y_2^p+...+y_n^p+\sum_{m=1}^{p-1} t_m (y_1^{[mr_1]_p}y_2^{[mr_2]_p}...y_n^{[mr_n]_p}) $$
where $[...]_p$ denotes mod $p$ value taking values $0\leq [...]_p<p$.  
Moreover we mod out this theory with a maximal subgroup $Z_p^{\otimes (n-1)}\subset Z_p^{\otimes n}$ which leaves the $W$
invariant.   The parameters $t_m$ denote the vevs of the $m$-th twisted sector chiral field.  If we consider product of abelian
orbifolds we get the same structure where for each twisted sector we get the associated deformations
as above.   Sometimes, as we will encounter later, the symmetries we mod out allow us
to define better variables $y_i^{k_i}\rightarrow y_i$ if all the monomials appearing in $W$ have $y_i$'s whose powers
are divisble by $k_i$.
\begin{table}
\begin{center}
\begin{tabular}{cccccccc}
$T^2/\Z_2 \colon$& \phantom{$\Big|$}$(0)_1$&$(1/2)_4$&$(1)_1$\\
$T^2/\Z_3\colon$ & \phantom{$\Big|$}$(0)_1$&$(1/3)_3$&$(2/3)_3$&$(1)_1$\\
$T^2/\Z_4\colon$& \phantom{$\Big|$}$(0)_1$&$(1/4)_2$&$(1/2)_3$&$(3/4)_2$&$(1)_1$\\
$T^2/\Z_6\colon$ &\phantom{$\Big|$}$(0)_1$&$(1/6)_1$&$(1/3)_2$&$(1/2)_2$&$(2/3)_2$&$(5/6)_1$&$(1)_1$\\
\end{tabular}
\caption{Dimensions and multiplicities of the allowed deformations of the LG mirrors of toroidal orbifolds $T^2/\Z_k$: the notation $(\ell/k)_{m_\ell}$ signify that there are $m_\ell$ fields with dimension $\ell/k$ in the chiral ring.}\label{toroidims}
\end{center}
\end{table}

Now consider the orbifold of $T^2$.  As shown in \cite{Vafa:2001ra} those of $T^2/(\Z_3,\Z_4,\Z_6)$ are given
by particularly simple LG models, namely
$$T^2/\Z_3:  \ \ W=x_1^3+x_2^3+x_3^3+ax_1x_2x_3+\text{defs}$$
$$T^2/\Z_4:  \ \ W=x_1^2+x_2^4+x_3^4+a x_1x_2x_3+\text{defs}$$
$$T^2/\Z_6: \ \ W=x_1^2+x_2^3+x_3^6+a x_ix_2x_3+\text{defs}$$
where $a$ parameterizes the complex structure of the mirror $T^2$ and the deformations involve
all the chiral fields in the LG model (see table \ref{toroidims}).  It is an easy exercise \cite{Vafa:2001ra} to check
that the geometry of the fixed point set of $T^2$ quotients match the chiral
deformations. An equally simple mirror for the $T^2/\Z_2$ is not available.  However a simple
way to obtain the mirror for a special class of these theories is to consider the case where $T^2$ complex
structure is $\tau =i$ (which will not appear in the ${\cal N}=2$ geometry in 4d) and start with the $T^2/\Z_4$ description above.  To obtain $T^2/\Z_2$
from this we can undo a $\Z_2$ which in the mirror is equivalent to modding out the theory
by a $\Z_2$:
$$T^2/\Z_2:  \ \ W=x_1^2+x_2^4+x_3^4+a x_1x_2x_3+\text{defs}/[(x_2,x_3)\rightarrow -(x_2,x_3)]$$
The chiral fields associated with the four fixed points of $T^2/\Z_2$ get mapped to $x_2^2,x_3^2,x_2x_3$ and the
twist field in this LG orbifold theory.\footnote{A more general $T^2/Z_2$ orbifold
may be obtained from the mirror of $T^2$ realization as a bidegree
polynomial in $\C\PP^1\times \C\PP^1$.  This leads, using the usual mirror symmetry arguments \cite{Hori:2000kt} to an LG theory with
$W=x_1^2+x_2^2+y+a x_1x_2+b x_1^2x_2^2/y$
where $x_i$ are $\C$ variables but $y$ is a $\C^*$ variable.}
For simplicity when we discuss the explicit ${\cal N}=2$ geometry we focus on the $\Z_3,\Z_4,\Z_6$ cases except
we also check the counting for the moduli also for the $\Z_2$ case, which only requires using the fact that there
are 4 dimension 1/2 chiral fields.

Now we combine the two ingredients:  We can mod out by a further symmetry so that the geometry takes
the form $T^2/\Z_p\times \C^2/\Z_p$.  This is clearly the tensor product of the two theories
which is simply the sum of the two W's.  Undoing the extra $\Z_p$ is equivalent (as is well known
in the context of mirror symmetry) to modding the two decoupled theories by an extra $\Z_p$.
As long as the holonomy is $SU(3)$ this amounts to writing all possible deformations $t_m$
made of the LG fields of the $T^2$ by requiring that the final $W$ is still quasi-homogeneous.
In other words we include all the combinations which are allowed by each sector of the orbifold and which
lead to a total charge 1 field in the superpotential.
This completes our quick review of mirror symmetry.  Before moving on to applications
we recall how this mirror can be used to construct the ${\cal N}=2$ vacuum geometry of the effective 4d theories.

\subsection{4d ${\cal N}=2$ geometry and Landau-Ginzburg geometry}
The mirror geometry generally leads to a Landau-Ginzburg theory \cite{Hori:2000kt} as discussed
in examples above.  Consider an LG theory with a quasi-homogeneous superpotential $W$. 
 Moreover we assume the LG theory is orbifolded  by $exp(iQ)$
which projects the theory to integral charge fields.
 We are particularly interested
in the case of the compactifications to 4 dimensions, which lead to a 2d ${\cal N}=(2,2)$ supersymmetric worldsheet theory with central charge $\hat c=3$, where 
$$\hat c=\sum_i(1-2q_i)$$
and $q_i$ is the charge of the corresponding fields (assigning weight 1 to $W$).  So for the $\C^*$ variables the $q_i=0$, as they are exponentiated fields in $W$. The reader can check for example in the $T^2\times \C^2$ orbifold examples discussed above $\hat c=3$ where $\hat c=1$ comes
from the three $x_i$ fields and $\hat c=2$ comes from the $y_1,y_2$.
To obtain the ${\cal N}=2$ vacuum geometry from a given LG theory, we have to reconstruct the BPS central charges of the 4d theory, which in the LG theory correspond to period integrals over non-trivial cycles $C_i$:
$$\int_{C_i} d\phi_i\ {\rm exp}(-W(\phi_i))$$
 In general there is no manifold description.  However as discussed in \cite{Greene:1988ut,Hori:2000kt} in the special case
where we have 5 fields this is equivalent to doing period integrals on the Calabi-Yau three-fold given by
$$W=0 $$
defined in the projectivization of the $\phi_i$-space with weights given by their charges.  This cuts the dimension down to three and it
defines a Calabi-Yau threefold geometry.  In this context the BPS central charges in 4d are simply the period integrals
of the holomorphic 3-form of the CY over the 3-cycles.

 However there are cases where $\hat c=3$ but the number of
variables $\phi_i$ is less than or more than 5.  If it is less than 5, the remedy is rather easy.  In that case we add additional
fields to the theory with quadratic superpotential which does not affect the IR theory,  and again end up with 5 fields, and get the CY geometry as above.
A particularly dominant example of this type, which arises in local mirror symmetry, is when we have only three fields
$y_i$ which are $\C^*$ variables.  In such a case adding the two quadratic fields, leads to a local CY 3-fold given by
$$W(y_1,y_2,y_3)-uv=0$$
going to a patch $y_3=1$ we get the equation
$$W(y_1,y_2,1)-uv=0$$
It is possible to show that in such a case the period integrals of the Calabi-Yau geometry reduces to integrals of the
1-form $\lambda= (log y_1) dy_2/y_2 $ over the curve
$$f(y_1,y_2)=W(y_1,y_2,1)=0$$
This is where we make contact with Seiberg-Witten geometry:  $f=0$ is the SW curve and $\lambda $ is the associated
1-form whose period integrals gives the vacuum geometry for the associated $4d$ theories with ${\cal N}=2$ supersymmetry.
However not all theories associated to interesting 4d ${\cal N}=2$ theories reduce to curves (see \cite{Katz:1997eq}) and we will encounter examples of this type here too.  

More interesting cases, which does arise in the mirror symmetry context, is when we have more than $5$ variables \cite{Vafa:1991uz}.  In such a case
we still can use the LG geometry to compute the period integrals.  But there is no associated Calabi-Yau 3-fold geometry or SW curve.
Nevertheless the LG geometry is sufficient to capture all the relevant vacuum geometry for the 4d ${\cal N}=2$ theory which is engineered as a non-geometric phase of type IIB.  We will also encounter examples of this
type later in this paper, when we consider reduction of $(T^2\times \C^2)/\mathsf{G}$ to 4d when $\mathsf{G}$ has certain non-abelian factors.

\subsection{Locating 4d conformal fixed points}
So far we have reviewed how the $W$ associated to a worldsheet superconformal theory captures the $4d$ vacuum geometry.
We have also explained how we can associate to it a Calabi-Yau geometry if the number of variables in $W$ is less than or equal to 5.
Let us first focus on this class.
We can write the resulting geometry in the form of a hypersurface (by going to a patch) as
$$f(z_1,z_2,z_3,z_4)=0$$
To obtain an ${\cal N}=2$ 4d SCFT conformal theory, we need to tune the mass and Coulomb branch parameters
so that the resulting CY geometry has singularities and moreover $f$ itself should be quasi-homogeneous \cite{Gukov:1999ya,Shapere:1999xr}.  Any 4d $\mathcal{N}=2$ SCFT has a $U(1)_R$ symmetry which means that the hypersurface $f=0$ has to have a scaling symmetry, a $\C^*$ action such that all the coordinates 
have positive weights:
\begin{equation}
f(\lambda^{q_i}z_i)=\lambda f(z_i),~~q_i>0;
\end{equation}
Treating this $f$ as a superpotential of a 2d theory we can associate a central charge.  The condition that
this theory is at finite distance in moduli space requires that $\hat c<2$  \cite{Gukov:1999ya,Shapere:1999xr}.  Such an $f$
can be viewed as a part of the original $W$ where we have an additional Liouville field which
makes up the balance of $\hat c$ from 3 \cite{Ooguri:1995wj}. 

Given an $f$ one can turn on deformations formed by monomials subject to the relations generated by 
\begin{equation}
{\partial f\over \partial z_i}=0,
\end{equation} 
$f$ plus all the deformations are just the ${\cal N}=2$ geometry. There is a canonical three form
\begin{equation}
\Omega={dz_1\wedge dz_2 \wedge dz_3\over \partial f},
\end{equation}
and we require $\Omega$ to have scaling dimension $1$ as the integration of this three form on three cycle would give the mass. For a deformation $u_{\alpha} z^{\alpha}$, one can find its scaling dimension 
\begin{equation}
[u]={2(1-Q_{\alpha})\over 2-\hat{c}},~~
\end{equation}
Here $Q_{\alpha}$ is the charge of the monomial $z^{\alpha}=z_0^a z_1^b z_2^c z_3^d$.  There are several simple comments \cite{Gukov:1999ya,Shapere:1999xr}:
\begin{itemize}
\item Only the deformation with non-negative scaling dimension can change the geometry, and the number of relevant deformations 
has to be finite, this means that $\hat{c}<2$. 
\item The deformations are paired except for the deformation with scaling dimension $1$, and they satisfy the following condition
\begin{equation}
[\lambda]+[u]=2.
\end{equation} 
This condition agrees with the scaling behavior of $\mathcal{N}=2$ supersymmetric relevant deformation, namely
they correspond to vev $[u]$ of an operator, or adding it to the prepotential with coefficient $[\lambda]$.
\end{itemize}

The spectrum of the positive deformations could be classified according to their scaling dimensions:
\begin{itemize}
\item $[u]> 1$: Coulomb branch operators.  These come from monomials with charge $Q_\alpha < {\hat c}/2$.  Among them, we say that an operator is relevant or exactly marginal respectively if its scaling dimension is $1<[u]<2$ of $[u]=2$;
\item The deformations with scaling dimension $[u]=1$ are called mass parameters, this happens if $Q_\alpha={\hat{c}\over2}$; 
\item The deformations with scaling dimension $ 0\leq[u]<1$ are called coupling constants.  These correspond to monomials with
 ${\hat c}/2<Q_\alpha \leq 1$.  An operator with $[u]=0$, corresponding to charge $Q_\alpha=1$, is an exactly marginal coupling. 
 \item The other monomials with $Q_\alpha >1$ are irrelevant deformations of the theory.
\end{itemize}

The singularity need not be isolated.  For example in the theories of class ${\cal S}$ we have a curve of $ADE$ type singularities which means that the above $f$ is of the form
$$f(z_1,z_2,z_3,z_4)=f_{ADE}(w_1(z_i),w_2(z_i),w_3(z_i))$$
where $f_{ADE}$ denotes the $ADE$ singularity.  $w_{1,2,3}=0$ gives a curve of singularities in the 4d ambient space of $z_i$.
The fact that $f_{ADE}$ are classified by the ADE follows from the fact that to reach a curve singularity at finite distance on moduli space
we need the singularity locus to be a quasi-homogenous polynomial with $\hat c <1$ instead of $\hat c <2$, because the dimension has
gone down by 1.   This gives an alternative derivation that the class ${\cal S}$ type theories are classified by $ADE$.
We will also encounter examples of the type where the singularity of the geometry is at infinity, where for example $f$ is of the form \cite{Xie:2012hs,Cecotti:2012jx}
$$f=exp(pY_0)+f_{ADE}(z_1,z_2,z_3)$$
If we have more than 5 variable we will not be able to get a geometric description, but again
we can tune parameters so that the associated $W$ even in a patch is quasi-homogeneous
singularity.  Such cases would lead to an $f$ which has more than 4 variables.  Nevertheless
for it to be reachable at finite distance again the associated $\hat c <2$.  An example of this type that
we will encounter is when
$$f=z_1^3+z_2^3+z_3^3 +f_{D,E}(z_4,z_5)$$
(where we have eliminated unnecessary quadratic fields from $f$).  
This can be viewed as coming from a $W$ given by
$$W=z_1^3+z_2^3+z_3^3+f_{D,E}(z_4,z_5)+\mu/z_6^{h}$$
where $h$ is the dual coveter number for the corresponding $D,E$ theory, and $\mu\not= 0$ is a deformation away
from the singular limit  \cite{Ooguri:1995wj}.  This $W$ will have $\hat c=3$. This gives novel examples
of 4d ${\cal N}=2$ theories where the vacuum geometry can be computed using period
integrals in an LG theory which we can obtain from 6d toroidal compactification and that we will
also identify directly in four dimensional terms.

\subsection{LG mirrors and the 5d theories of ${\hat D}_p(G)$ type}

As a first application of the mirror symmetry methods we just described it is possible to check on the proposal that the 5d ${\hat D}_p(G)$ do indeed lead to $D_p(G)$ theories upon reduction on a circle. For simplicity let us consider the case $\Gamma = \Z_n$. We are considering an orbifold generated by two elements, namely
\be
\mathsf{G} \equiv\langle g = (\alpha^{2},\alpha^{-1},\alpha^{-1}) , h = (1,\omega,\omega^{-1}) \rangle \text{ with } \alpha \in \Z_{2p} \quad \omega \in \Z_n.
\ee
The mirror of this system is readily obtained with the methods illustrated in the previous section. We have
\be\label{5dpGxxxx}
W = \begin{cases}
&\sum_{i=1}^3 y_i^{np} + \sum_{r,m} t_{r,m} (y_1^{[ n r ]_{np}} y_2^{[- \frac{n}{2} r +p m]_{np}} y_3^{[-\frac{n}{2} r - p m]_{np}}) \qquad (n \text{ even} )\\
&\sum_{i=1}^3 y_i^{2np}+ \sum_{r,m} t_{r,m} (y_1^{[2 n r]_{2np}} y_2^{[- n r + 2 p m]_{2np}} y_3^{[- n r  - 2 p m]_{2np}}) \qquad (n \text{ odd} )
\end{cases}
\ee
where the sum is taken over the set of deformations with dimension 1 and all the $y_i$ are $\C^*$ variables. By construction, we can always tune the deformation parameters corresponding to the coefficients without a $y_1$ term to be of the form 
$$(y_2^p+y_3^p)^n \quad (n \text{ even }), \qquad (y_2^{2p}+y_3^{2p})^n \quad (n \text{ odd }),$$ 
moreover, notice that there is an extra $\Z_n$ symmetry (resp. $\Z_{2n}$ symmetry) as $y_1$ always appears to the $n$-th power (resp. to the $2n$-th power) which we have to mod out.  This can be accomplished by changing to a new variable $y = y_1^n$ (resp. $y=y_1^{2n}$) which is still a $\C^*$ variable. Moreover, tuning to zero all the $t_{r,m}$ coefficients of all monomials with $y_1 \neq 0$ we can rewrite the LG superpotential of eqn.\eqref{5dpGxxxx} as follows:
\be
W = y^p + (y_2^p+y_3^p)^n \quad (n \text{ even }), \qquad W = y^p +(y_2^{2p}+y_3^{2p})^n \quad (n \text{ odd })
\ee
Specializing to the patch $y_3 = 1$ gives a $\C$ variable $w = (y_2^p +1)$ (resp. $w = (y_2^{2p} +1)$) and we end up with the geometry
\be
uv = f(y,w) = y^p + w^n 
\ee
which is exactly the SW geometry for $D_p(SU(n))$ \cite{Xie:2012hs,Cecotti:2012jx,Cecotti:2013lda}. Proceeding analogously, replacing $\Z_n$ with $\Gamma$ a $DE$ subgroup of $SU(2)$, one obtains
\be
y^p + f_G(w,u,v) = 0
\ee
where $f_G$ is an $ADE$ singularity. This is the $D_p(G)$ geometry which arises from the decoupling limit considered in \cite{Cecotti:2012jx,Cecotti:2013lda}.

\section{$SL(2,\Z)$ duality from 6d:  $(E_n^{(1,1)},G)$ models}

In this section we cross check our findings of \S.\ref{mtheoryengine} using the mirror geometries. The 4d theories we are going to identify have an exactly marginal deformation which corresponds to the complex structure of the torus, which predicts that these systems enjoy an exact $SL(2,\Z)$ duality symmetry. Due to our geometric construction, all these theories are self-dual, and are natural generalizations of four dimensional  $\mathcal{N}=4$ SYM, whose $SL(2,\Z)$ invariance is best understood from compactifying 6d $(2,0)$ theory on $T^2$.

\subsection{Mirror geometries of $\cO(-n)$ models}

We begin by considering the simple case of orbifold models identified in \S.\ref{originalorbfs}. Notice that the $\Z_2$ element of that class of models is simply the six dimensional $A_1$ (2,0) theory and it is well known that we get four dimensional $\mathcal{N}=4$ SYM by compactifying it on $T^2$. The gauge coupling of the 4d theory is interpreted as the complex structure of the torus, and the S duality group is interpreted as the mapping class group of the torus. The superconformal invariance of 6d theory play a crucial role in deriving 4d S duality \cite{Witten:2009at}. Here we focus on the remaining five cases, namely $\Z_n$ with $n={3,4,6,8,12}$.   Note that for $n\not=3$ we still will have a $Z_2$ sector, and thus
an $A_1$ singularity.  So, as discussed before, we expect to get for these theories an $SU(2)$ gauge theory coupled to matter when we
compactify down to 4 dimensions.  For $n=3$, the story will be different.

\begin{table}
\begin{center}
  \begin{tabular}{cccccc }
   \phantom{$\Bigg|$  } $\cO(-3)$ & $\cO(-4)$ &$\cO(-6)$& $\cO(-8)$&$\cO(-12)$ \\ 
   \hline
  \phantom{$\Bigg|$  }4 &6 &8&9&10 \\ 
  \end{tabular}
      \caption{The number of parameters for the $T^2$  compactification of minimal 6d SCFT.}
  \label{mini}
\end{center}
\end{table}

One can write down the corresponding $\cN=2$ geometry using mirror symmetry as discussed in section 4 (for this class of models see also \cite{Haghighat:2014vxa}, in particular, the LG mirror of the $\cO(-3)$ model presented here was worked out there).   The mirror theory has two parts: the $T^2$ part is described by three $\C$ variables $x_1, x_2, x_3$, and 
the $\C^2$ part described by two $\C^*$ variables $y_1, y_2$. The two $y$ variables for the present models both have charge $1$ under the orbifold action, so the allowed $y$ monomials have the form $(y_1y_2)^i$. The deformations of the corresponding LG mirrors are obtained by the weight one monomials built out of the products of the allowed $y$ monomials and the $x$ variables. We obtain the following LG mirrors for the $T^2$ compactifications:
\be
\begin{aligned}
&  W_{\cO(-3)}=~~ W_{T^2/\Z_3}+ y_1^3+y_2^3+ y_1y_2(\sum_i \beta_i x_i); \\
&  W_{\cO(-4)}=~~W_{T^2/\Z_2} + y_1^4+y_2^4+y_1y_2(4 \ {\rm dim}\ 1/2 \ {\rm twist fields})+ \beta_5 y_1^2y_2^2;  \\
&   W_{\cO(-6)}=~~W_{T^2/\Z_3}+ y_1^6+y_2^6+ y_1y_2(\b_1 x_1 x_2+ \b_2 x_1x_3+ \b_3 x_2 x_3)+  \\
&~~~~~~~~~~~~~y_1^2y_2^2(\b_4 x_1+\b_5x_2+\b_6 x_3)+\b_7 y_1^3y_2^3; \\
&   W_{\cO(-8)}=~~W_{T^2/\Z_4}+y_1^8+y_2^8+y_1y_2(\beta_1 x_2^2x_3+ \beta_2 x_3^2x_2)+  \\
&~~~~~~~~~~~~~y_1^2 y_2^2(\beta_3 x_2^2+\beta_4 x_3^2+\beta_5 x_2x_3)+y_1^3y_2^3(\beta_6 x_2+\beta_7 x_3)+\beta_8 y_1^4y_2^4;  \\
&   W_{\cO(-12)}~~W_{T^2/\Z_6}+y_1^{12}+y_2^{12}+y_1y_2(\beta_1 x_2x_3^3)+y_1^2y_2^2(\beta_2 x_2 x_3^2+\beta_3 x_3^4) \\
& ~~~~~~~~~~~~~y_1^3y_2^3(\beta_4 x_2 x_3+\beta_5 x_3^3)+y_1^4y_2^4(\beta_6 x_2+ \beta_7x_3^2)+\beta_8 y_1^5y_2^5 x_3+\beta_9 y_1^6y_2^6.
\end{aligned}
\ee
 It is easy to check that we get the same number of parameters found from the 6d description in table \ref{mini}. 

\begin{center}
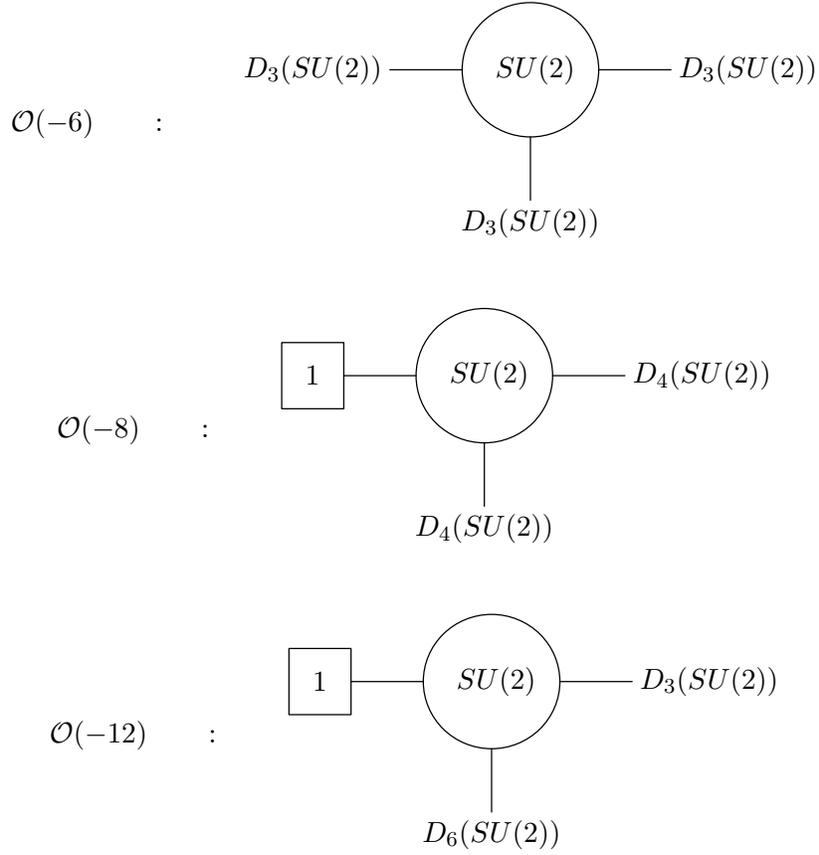
\begin{figure}
\small
\centering

$$\cO(-6)\qquad \colon \qquad\begin{gathered}\xymatrix{D_3(SU(2))\ar@{-}[r]&*+++[o][F-]{\phantom{\Bigg|}SU(2)} &D_3(SU(2))\ar@{-}[l]\\&D_3(SU(2))\ar@{-}[u]&}\end{gathered}$$

$$\cO(-8)\qquad \colon \qquad\begin{gathered}\xymatrix{*+++[][F-]{1}\ar@{-}[r]&*+++[o][F-]{\phantom{\Bigg|}SU(2)} &D_4(SU(2))\ar@{-}[l]\\&D_4(SU(2))\ar@{-}[u]&}\end{gathered}$$

$$\cO(-12)\qquad \colon \qquad\begin{gathered}\xymatrix{*+++[][F-]{1}\ar@{-}[r]&*+++[o][F-]{\phantom{\Bigg|}SU(2)} &D_3(SU(2))\ar@{-}[l]\\&D_6(SU(2))\ar@{-}[u]&}\end{gathered}$$

\caption{4d SCFT found from compactifying 6d minimal SCFT on $T^2$. here $D_p(SU(2))$ is the $(A_1, D_p)$ Argyres-Douglas theory found in \cite{Eguchi:1996vu}.}
\label{Su2}
\end{figure}
\end{center}

As discussed in the beginning of this section, the complex structure of the torus has to  
be the exactly marginal deformation of the 4d theory: tuning the parameters to reach a superconformal point, we are not going to alter the $T^2$ part of the corresponding LG model. 
 This type of singular geometry can be found explicitly for the $\cO(-n)$ models with $n=6,8,12$.  Motivated
 by the connection with $SU(2)$ gauge theory, to find a singular point we need to go to the origin
 of Coulomb branch, which geometrically is the $A_1$ singularity in the geometry.  This we can do by simply tuning 
the Coulomb branch parameters involving $y$'s only to get a quadratic term.  We obtain: 
\be
\begin{aligned}
& W_{\cO(-6)}=~~x_1^3+x_2^3+x_3^3+\alpha \,x_1x_2x_3+(y_1^3+y_2^3)^2; \\
& W_{\cO(-8)}=~~x_1^2+x_2^4+x_3^4+\alpha \, x_1x_2x_3+(y_1^4+y_2^4)^2; \\
& W_{\cO(-12)}=~~x_1^2+x_2^3+x_3^6+\alpha \, x_1x_2x_3+(y_1^{6}+y_2^6)^2. 
\end{aligned}
\ee
By going to an affine patch $y_2=1$ and introducing a new $\C$ variable $w = y_1^k+1$:
\be\label{singular}
\begin{aligned}
& f_{\cO(-6)}=~~x_1^3+x_2^3+x_3^3+\alpha \, x_1x_2x_3+w^2=0; \\
& f_{\cO(-8)}=~~x_1^2+x_2^4+x_3^4+\alpha \, x_1x_2x_3+w^2=0; \\
& f_{\cO(-12)}=~~x_1^2+x_2^3+x_3^6+\alpha \, x_1x_2x_3+w^2=0. 
\end{aligned}
\ee
These singular geometries give the elliptic 4d SCFTs of \cite{Cecotti:2011rv}.  These theories have
 weakly coupled gauge theory descriptions as an $SU(2)$ gauge group weakly gauging the diagonal flavor symmetry of some $D_p(SU(2)$
 matter systems, see fig.\ref{Su2}. This confirms our findings of section \ref{mtheoryengine}. These models are labeled by $D_4^{(1,1)}, E_6^{(1,1)}, E_7^{(1,1)}$ and $E_8^{(1,1)}$, as the charges of the corresponding BPS particles belong to the root lattices of the corresponding extended affine Lie algebras \cite{Cecotti:2012va}. In particular, the fact that these algebras are doubly extended gives rise to a natural $SL(2,\Z)$ action on the set of imaginary roots, which was identified with the $S$--duality group in \cite{Cecotti:2012va}. We have just found the geometric 6d origin of such an exact $SL(2,\Z)$ $S$--duality!

\begin{table}
\begin{center}
  \begin{tabular}{c | cc c c c }
  
    ~ & Coulomb & mass  & marginal & relevant & total  \\ \hline
   $D_4^{1,1}$ \phantom{$\Bigg|$}& 1 & 4 &1&0&6\\ \hline
   $E_6^{1,1}$ \phantom{$\Bigg|$}& 4 & 0 &1&3& 8\\ \hline
   $E_7^{1,1}$ \phantom{$\Bigg|$}& 3 & 3 &1&2& 9\\ \hline
   $E_8^{1,1}$ \phantom{$\Bigg|$}& 4 & 2 &1&3&10\\ 
    \hline
  \end{tabular}
  
  \caption{The number of physical parameters of 4d SCFTs which are found at the most singular point of $T^2$ compactification of 6d minimal SCFT.}\label{wroirwrwoib}
\end{center}
\end{table}


The gauge coupling of each elliptic 4d theory gets identified with the complex structure of the torus, and the S-duality group of the 4d theory is identified with the mapping class group of it: this implies that such theories are self-dual. This fact was noticed for the $E_7^{(1,1)}$ case in \cite{Buican:2014hfa} and indeed the theory is self-dual under the $SL(2,\Z)$ duality symmetry, in perfect accord with our findings.

Let us conclude this section by considering the  $\Z_3$ case. In this case we were not able to find a locus of the Coulomb branch moduli where an exactly marginal coupling emerges.   This is consistent with the fact that we do not have an $A_1$ singularity
to lead to weakly coupled $SU(2)$ gauge system.  We were however able to locate a $D_4$ AD point. This is discussed in appendix \ref{D4singularata}.

\subsection{Conformal matter and $(E_n^{(1,1)},G)$ theories}

\begin{figure}
\begin{center}
\small
\centering

$$(D_4^{(1,1)},SU(2m))\quad\colon\quad\begin{gathered}\xymatrix{
*+++[o][F-]{m}\ar@{-}[dr]&&*+++[o][F-]{m}\ar@{-}[dl]\\
&*+++[o][F-]{2m}\\
*+++[o][F-]{m}\ar@{-}[ur]&&*+++[o][F-]{m}\ar@{-}[ul]\\}\end{gathered}$$

$$(E_6^{(1,1)},SU(3m))\quad\colon\quad\begin{gathered}\xymatrix{
&&*+++[o][F-]{m}\ar@{-}[d]\\
&&*+++[o][F-]{2m}\ar@{-}[d]\\
*+++[o][F-]{m}\ar@{-}[r]&*+++[o][F-]{2m}\ar@{-}[r]&*+++[o][F-]{3m}\ar@{-}[r]&*+++[o][F-]{2m}\ar@{-}[r]&*+++[o][F-]{m}}\end{gathered}$$

$$(E_7^{(1,1)},SU(4m))\quad\colon\quad\begin{gathered}\xymatrix{
&&&*+++[o][F-]{m}\ar@{-}[d]\\
&&&*+++[o][F-]{2m}\ar@{-}[d]\\
*+++[o][F-]{m}\ar@{-}[r]&*+++[o][F-]{2m}\ar@{-}[r]&*+++[o][F-]{3m}\ar@{-}[r]&*+++[o][F-]{4m}\ar@{-}[r]&*+++[o][F-]{3m}\ar@{-}[r]&*+++[o][F-]{2m}\ar@{-}[r]&*+++[o][F-]{m}}\end{gathered}$$

$$(E_8^{(1,1)},SU(6m))\quad\colon\quad\begin{gathered}\xymatrix{
&&&&&*+++[o][F-]{2m}\ar@{-}[d]\\
&&&&&*+++[o][F-]{4m}\ar@{-}[d]\\
*+++[o][F-]{m}\ar@{-}[r]&*+++[o][F-]{2m}\ar@{-}[r]&*+++[o][F-]{3m}\ar@{-}[r]&*+++[o][F-]{4m}\ar@{-}[r]&*+++[o][F-]{5m}\ar@{-}[r]&*+++[o][F-]{6m}\ar@{-}[r]&*+++[o][F-]{3m}}\end{gathered}$$

\caption{Affine quiver which are also in our $(E_n^{(1,1)}, SU(N))$ list.}
\label{affine}
\end{center}
\end{figure}

In this section we will generalize the above construction to obtain arbitrary ADE group $G$ generalizing from the $A_1$ case discussed
above.  To do this, let us first consider the case of the $\cT(E_6,N)$ theory, the worldvolume theory of a stack of $N$ M5 branes probing the $E_6$ singularity. We can read off the LG mirror geometry directly in terms of the orbifold action of eqn.\eqref{conformattorbofo}. The LG description of the IIB mirror CY is

\be\label{confLGe6}
\begin{aligned}
W_{\cT(E_6,N)}(x_1,x_2,x_3,y_1,y_2)= & \frac{x_1^3}{3}+\frac{x_2^3}{3} + \frac{x_3^3}{3} + \frac{y_1^{3N}}{3N} +\frac{y_2^{3N}}{3N} \\
& + \sum_{i=1}^3\sum_{a+b = N} t^{(1)}_{i,a,b} \,x_i \, y_1^a  \, y_2^b  +  \sum_{i=1}^3\sum_{a+b = 2N} t^{(2)}_{i,a,b} \, x_i^2\,  y_1^a  \, y_2^b \\
& +  \a \, x_1 x_2 x_3 + \sum_{a+b = 3N\atop a,b\neq 3N}  t^{(3)}_{a,b} \, y_1^a  \, y_2^b
\end{aligned}
\ee

where the $t^{(j)}$'s are deformation parameters, while $\a$ corresponds to the size of the $T^2$. Notice that in this case the $\C^*$ variables $y_i$ are unconstrained: the only constraint is that these have to match the deformations of the $T^2/\Z_3$ mirror. This LG geometry has precisely 
\be
1+3(N+1) + 3(2N+1)+ (3N - 1) = 12N + 6 
\ee
parameters, which, from the description given in section \ref{orbfsandconfmat}, precisely equals $L(\cT(E_6,G))$. To get the corresponding CY, we proceed in the standard way and we consider a special patch, for example $y_2=1$. From the explicit expression, it is clear that we can tune the deformation parameters setting $t^{(1)}_{i,a,b} = t^{(2)}_{i,a,b}=0$ in such a way that we obtain the following CY hypersurface:
\be
0= \frac{x_1^3}{3}+\frac{x_2^3}{3} + \frac{x_3^3}{3} + \a x_1 x_2 x_3 + (y_1+1)^{3N}
\ee
Notice that we can trade the $\C^*$ variable $y_1$ for a $\C$ variable $w=y_1+1$, and in this way we obtain an isolated CY singularity of the form
\be
0= \frac{x_1^3}{3}+\frac{x_2^3}{3} + \frac{x_3^3}{3} + \a x_1 x_2 x_3 +w^{3N}
\ee
Such singularity corresponds to a LG model with $\hat{c}<2$ and is therefore at finite distance in CY moduli space. This singularity indeed corresponds to the theory $(E_6^{(1,1)},SU(3N))$ \cite{Katz:1997eq}:\footnote{ Our readers which are not familiar with such systems can find a detailed summary of most of the relevant properties for the models of interest in this paper in appendix \ref{ellipticSCFTs}.} we obtain an affine $E_6$ quiver theory of the type in figure \ref{affine} with $m=N$.

The other cases are analogous. The LG mirrors are given by
\be\label{LGconformatta}
W = W_{T^2/\Z_r}(x_i) + y_1^{kN} +y_2^{kN} + \sum_{\ell=0}^k\sum_{j=1}^{m_\ell} \sum_{a+b = \ell N\atop a,b\neq kN} t_{\ell,j,a,b} \, \varphi_{j,\ell/k}(x_i) \, y_1^a \, y_2^b
\ee
where $\varphi_{j,\ell/k}(x_i)$ denotes a chiral ring element of dimension $\ell/k$, and $j=1,...,m_\ell$ denotes the corresponding multiplicity we summarized in table \ref{toroidims}. From the above equation, these systems have
\be\label{paramatter}
\begin{aligned}
& k = 2 \colon 1+4(N+1) + (2N -1) = 6N+4 \\
& k = 4 \colon 1+ 2 (N+1) + 3 (2N+1) + 2 (3N+1)+(4N-1)=18N + 7\\
& k = 6 \colon 1 +(N+1) + 2(2N+1) + 2(3N+1) + 2(4N+1) + (5N+1) + (6N-1)\\
&\hspace{12cm}= 30N + 8\\
\end{aligned}
\ee
parameters which matches with the various $L$'s for the conformal matter systems as computed from section \ref{orbfsandconfmat}. Again, it is easy to see that in the patch $y_2=1$ we can turn off all $t_{\ell,j,a,b}$ with $\ell\neq 0$ and, by fine tuning the $t_{0,1,a,b}$ coefficients and setting $w=y_1+1$ obtain the hypersurface singularity
\be
0 = W_{T^2/\Z_k}(x_i)  + w^{kN}
\ee
which, for $k=2,3,4,6$ respectively corresponds to a conformal affine quiver theory of type $D_4$, $E_6$,$E_7$ and $E_8$ in figure \ref{affine} with $m=N$.  So we have obtained a subset of theories $(E_n^{(1,1)},A_{r-1})$ for which $r$ has divisors (2,3,4,6) for $n=4,6,7,8$ respectively from compactification from 6d. In the next section we continue with the more general case. In these cases the $D_{p_i}(SU(r))$ theories involved are all Lagrangian (see appendix \ref{ellipticSCFTs}). Note that the fact that the moduli space of these theories are given
by flat $ADE$ connections on $T^2$ \cite{Katz:1997eq} has now found a natural 6d interpretation.  See also \cite{Nekrasov:2012xe,Nekrasov:2013xda} for
a study of the Nekrasov partition function for this class of theories.

\subsection{$\langle \Z_{4,6,8,12}, \Gamma_{ADE}\rangle $ (1,0) 6d theories on $T^2$ and 4d $(E_n^{(1,1)},G_{ADE})$}
In the previous section we have argued how we can obtain 4d theories of $(E_n^{(1,1)},A_{r-1})$ type for some
$r$'s.  Here we show more generally how we can get all the theories in 4d of the type $(E^{(1,1)},G_{ADE})$.  In fact
as discussed in section 3, we expect the orbifold 6d SCFT's where the orbifold group is given by $\Z_{4,6,8,12}\times \Gamma_{ADE}$
(modulo a $\Z_2$ action if $ADE$ includes the center of $SU(2)$) should lead to the corresponding theory in 4d.  Recall
that $\Z_{4,6,8,12}$  is composed of $\Z_{2,3,4,6}$ action on the $T^2$ and rotation of the $\C^2$ coordinates
by $\Z_{4,6,8,12}$ and $\Gamma_{ADE}$ acts purely on the $\C^2$ part.  From this description, as we have discussed it
is clear that we can get an $ADE$ gauge symmetry in 4d as in the usual ADE (2,0) theories.  Moreover as we have argued
the $\Z_{4,6,8,12}$ generated leads to certain matters of the type $D_p(G_{ADE})$ for each fixed point
of $T^2/\Z_{2,3,4,6}$ where $p$ is the order of the stabilizer of the fixed point.  We now use mirror symmetry to compute the resulting
${\cal N}=2$ geometry which can be equivalently be viewed as a way to solve for the vacuum geometry of the $(E_n^{(1,1)},G_{ADE})$ theories.
We first focus on the $A$ case and explain how it generalizes to the other cases.

We start with the mirror for $T^2/\Z_k$, which we denote by $W_{T^2/\Z_k}$.  We then add the mirror associated to $\C^2$ orbifold.  
Let us consider the case $A_{N-1}$.  
For simplicity of presentation let us assume $N$ and $k$ are relatively prime (though the generalization
to other cases is straight-forward). The orbifold is given by
$$W=W_{T^2/\Z_k}+y_1^{Nk}+y_2^{Nk}+{\rm deformations}$$
where the deformations can include terms which mix the two parts, coming from the sectors where
the orbifold action is non-trivial on both $\C^2$ and $T^2$ or non-mixed part, coming from the sectors where
the action on $T^2$ is trivial.  Since $\Gamma_{A_{N-1}}$ is of this latter type, it means that we will get unmixed deformations
$$y_1^{ik}y_2^{Nk-ik}$$
which are the only ones we will use.  In particular since we want to go to the origin of the Coulomb branch for the $A_{N-1}$ gauge
theory in 4d, this means we want to be at the singular locus of the $A_{N-1}$ geometry, which means we have to turn on deformations
which lead to the singularity:
$$W=W_{T^2/\Z_k}+(y_1^k+y_2^k)^N$$
Going to the $y_2=1$ patch and redefining $w=y_1^k+1$, this gives us a CY geometry of the form
$$f=W_{T^2/\Z_k}+w^N=0$$
This, up to deformations is the vacuum CY geometry for the $(E_n^{1,1},A_{N-1})$.  As is clear from this argument
the mixed sectors of the orbifold do not participate in getting the geometry, and so this construction generalizes to the full ADE case.
In $D,E$ cases we can use the fact that mirror of $D,E$ are again given by $LG$ theories as in \cite{Ooguri:1995wj} to come up with the superpotential
$$W=W_{T^2/\Z_k}+W_{ADE}(z_1,z_2,z_3; z)$$
where $z$ is a Liouville field.  For all $(E_n^{(1,1)},G_{A,D,E})$ except $(E_6^{(1,1)},G_{D,E})$, by getting rid of unnecessary quadratic terms in the above, we can
get the geometry of a Calabi-Yau 3-fold (by going to the $z=1$ patch) of the form
$$f=W_{T^2/\Z_{4,6}}(x_1,x_2)+W_{ADE}(z_1,z_2)=0$$
which leads to the ${\cal N}=2$ vacuum geometry.  For $E_6^{(1,1)}$ case we can do the same for the $A$ case:
$$f=x_1^3+x_2^3+x_3^3+ax_1x_2x_3+W_{A}(z_1)=0$$
But for $D,E$ we only have
the LG form for $E_6^{(1,1)}$:
$$W=x_1^3+x_2^3+x_3^3+ax_1x_2x_3+W_{DE}(z_1,z_2;z).$$
This is not related to a local Calabi-Yau threefold, but it still can be used to compute the ${\cal N}=2$ vacuum geometry.

\section{Class $\cS$ from mirror geometry}

In the last section, we have successfully located one type of 4d SCFT from compactifying 6d theory on $T^2$.  For these 4d theories, 
which can be viewed as 6d $(1,0)$ theories obtained from orbifolds of $(2,0)$ ADE theories, compactified on $T^2$, the $ADE$  gauge symmetry emerges in 4d as is usual, where its coupling is identified with 
the complex structure of the torus, and the resulting ${\cal N}=2$ theory inherits the $SL(2,\Z)$ duality symmetry of $T^2$ as in the
$(2,0)$ case. From the $\cN=2$ geometry point of view, we did not touch the geometry 
associated with the $T^2$ part in locating the 4d SCFT. In this section, we are going to locate a different kind of  4d SCFT by tuning the parameter involving 
the $T^2$ part which is motivated by turning off certain flavor Wilson lines, as discussed in section 3. 
In the process, we find an emerging punctured Riemann surface which appears also in the $(2,0)$ compactification, and 
the S duality is interpreted as the mapping class group of this emerging punctured Riemann surface. Using our $\cN=2$ geometry, it is possible for us to determine the puncture type.  We will mostly be discussing the case of $(G,G)$ and $(G,G')$ conformal matter which lead
to certain genus $0$ class ${\cal S}$ theories, but also comment on obtaining some higher genus class ${\cal S}$ theories
along the lines of what we discussed in section 2 for $A$-type $(1,0)$ theories in 6d.

\subsection{$(E_6,E_6)$ conformal matter}
The LG model of eqn.\eqref{confLGe6} can be rewritten as follows
\begin{equation}
W = x_1^3+x_2^2x_3+x_3^3+ax_1x_2 x_3 +y_1^{3N}+y_2^{3N}+\sum f_i(y)g_i(x),
\end{equation}
where for later convenience the curve for the $T^2$ part is chosen in a slightly different but equivalent form. The orbifold action implies that the allowed $y$ monomial has the form $y_1^i y_2^j$ with $i+j=Np,~~p=0,1,2,3$. 
The deformations are chosen so that each monomial $f_i(y)g_i(x)$ has weight one. By tuning the parameters, we can re-cast $W$ in the following interesting form
\begin{align}
&W = x_1^3+x_2^2 g(x_3,y_1,y_2)+g(x_3,y_1,y_2)^2x_3+g(x_3,y_1,y_2)^3; \nonumber\\
& g(x_3, y_1,  y_2)= x_3+y_1^N+y_2^N+\sum_{i=1}^{N-1} a_i y_1^iy_2^{N-i}.
\end{align}
Naively, one would like to keep the $g(x_3,y_1,y_2)^3$ term as the most singular one, however, keeping only this term gives us a singularity with $\hat{c}=2$ which cannot correspond to a 4d SCFT as it is not at finite distance in moduli space. So the most singular geometry for a 4d SCFT is found by keeping the first three terms and dropping the last one, as it becomes irrelevant. Then, we go to affine patch $x_3\neq0$, and use the scale invariance to set $x_3=1$. We obtain the following local three--fold:
\be
\begin{cases}
&f = x_1^3+x_2^2\rho+\rho^2=0, \\
& \rho=g(1,y_1,y_2)=1+y_1^N+y_2^N+\sum_{i=1}^{N-1} a_i y_1^iy_2^{N-i};
\label{e6conformal}
\end{cases}
\ee
The above geometry means that there is an $E_6$ singularity over the curve $\rho=0$ which defines an $N+2$ punctured sphere. Therefore, we find 
an emerging punctured Riemann surface.  Here $a_i$ are the $N-1$ parameters which control the complex structure of the punctured Riemann surface which is identified 
with the exact marginal deformations of our 4d SCFT.  Surprisingly, starting with a 6d $(1,0)$ SCFT and compactifying on $T^2$, we can directly get a  class ${\cal S}$ description  in which the curve emerges using mirror symmetry.  We can also directly identify the details of the class ${\cal S}$ description, namely 
the puncture type  using our geometry, as we will now show.

The punctured Riemann surface defined by $g(1,y_1, y_2)=0$ has two distinguished punctures at $y_1=0$ or $y_2=0$. 
We now prove that these punctures  are $E_6$ full punctures of class ${\cal S}$ construction, and we also prove that the other $N$ punctures are simple. 
To simplify the notation, we take $N=1$, then $g=1+y_1+y_2=0$ defines
a three punctured sphere, and three punctures are $y_1=0$, $y_1=-1$ (equivalently $y_2=0$), and $y_1=\infty$.  Using $y_1$ as the coordinate
for the punctured sphere, the  $\cN=2$ geometry with all deformation terms is:
\be
 \begin{cases}
 &f=w^2+x_1^3+x_2^2\rho+\rho^2+(m_1+m_1^{'} y_1 )x_1x_2^2+(m_2+m_2^{'} y_1) x_1x_2+  (m_3 +u_1 y_1+m_3^{'}y_1^2) x_2^2 \nonumber\\
 & ~~~+ (m_4+u_2 y_1+m_4^{'} y_1^2) x_1+(m_5+u_3 y_1+m_5^{'}y_1^2) x_2+(m_6+u_4y_1+u_5y_1^2+m_6^{'}y_1^3)=0  \nonumber\\
& \rho=(1+y_1+y_2).
\end{cases}
\label{4dcurveBBB}
\ee
The deformations are found from \eqref{e6conformal} by setting $x_3=1$, and we also use the following conditions: 
 $x_2^2=\rho$ due to the ring relation of the singular geometry. We organize the curve in this form so 
it can be directly compared with the curve of class ${\cal S}$ construction.
 The total number of deformations is $17$ which is one less than for the 6d theory on $T^2$. 
In this case the torus complex structure deformation term $x_1x_2x_3$ is no longer an exactly marginal deformation.

The holomorphic top-form is $\Omega={dx_1 \wedge dx_2 \over w} \wedge {dy_1  \over y_1}$. Let's make a further change of coordinates 
\begin{equation}
w=\tilde{w}y_1^6,~~x_1=\tilde{x_1} y_1^4,~~x_2=\tilde{x_2} y_1^3,
\label{change}
\end{equation} 
then the above geometry becomes:
\be
 \begin{cases}
 &f=\tilde{w}^2+\tilde{x_1}^3+\tilde{x_2}^2\rho+\rho^2+({m_1\over y_1^2}+{m_1^{'}\over y_1} )\tilde{x_1}\tilde{x_2}^2+({m_2\over y_1^5}+{m_2^{'}\over y_1^4}) \tilde{x_1}\tilde{x_2}\\
&~~  +({m_3\over y_1^6}+{u_1\over y_1^5}+{m_3^{'}\over y_1^4}) \tilde{x_2}^2 +({m_4\over y_1^8}+{u_2\over y_1^7}+{m_4^{'}\over y_1^6}) \tilde{x_1}  +({m_5\over y_1^9}+{u_3\over y_1^8}+{m_5^{'}\over y_1^7} )\tilde{x_2} \\
&~~  +({m_6\over y_1^{12}}+{u_4\over y_1^{11}}+{m_6^{'}\over y_1^{9}})=0; \\
&\rho=(1/y_1^6+1/y_1^5+y_2/y_1^6).
 \label{4dcurve}
 \end{cases}
\ee
The holomorphic 3-form is now $\Omega={d\tilde{x_1} \wedge d\tilde{x_2} \wedge d y_1\over \tilde{w}}$, and $\tilde{w}\in K^6,~\tilde{x_1}\in K^4,~\tilde{x_2}\in K^3$ with $K$ the 
canonical bundle on Riemann surface parameterized by $y_1$. In the above expression, mass and Coulomb branch deformations are encoded in the terms in parenthesis: the leading order pole 
gives the mass deformation, and the subleading gives the contribution to Coulomb branch. The pole 
structure of this puncture is $(1,4,5,7,8,11)$.  The same analysis can be applied to the puncture at $y_2=0$, and we conclude that the puncture type is the same. 
 
Let us now analyze the behavior near the puncture $y_1= \infty$. To analyze the singular behavior of our geometry near this point, we change to the coordinate $y_1^{'}={1\over y_1}$ and obtain
\be
\begin{cases}
 &f = \tilde{w}^2+\tilde{x_1}^3+\tilde{x_2}^2\rho+\rho^2+({m_1^{'}\over y_1^{'}} )\tilde{x_1}\tilde{x_2}^2+({m_2^{'}\over y_1^{'}}) \tilde{x_1}\tilde{x_2}\\
 &~~+({m_3^{'}\over y_1^{'2}}) \tilde{x_2}^2 +({m_4^{'}\over y_1^{'2}}) \tilde{x_1}  
 +({m_5^{'}\over y_1^{'2}} )\tilde{x_2}+({m_6^{'}\over y_1^{3'}})=0\\
&\rho=(1/y_1^{'6}+1/y_1^{'5}+y_2/y_1^{'6}).
 \label{6d}
 \end{cases}
 \ee
So the order of pole near the puncture $y_1=\infty$ is $(1,1,2,2,2,3)$. In fact, these numbers are just the highest exponent of $y_1$ in \eqref{4dcurve}.

The $\cN=2$ geometry of a class ${\cal S}[E_6]$ theory can be  written in the following form \cite{Chacaltana:2014jba}:
\begin{equation}
w^2+x^3+y^4+\epsilon_2(z) x y^2+\epsilon_5(z) xy +\epsilon_6(z) y^2+\epsilon_8(z) x+\epsilon_9(z) y+\epsilon_{12}(z)=0;
\label{e6curve}
\end{equation}
here $\epsilon_i(z)$ is the degree $i$ differential on the Riemann surface parameterized by $z$, and $w\in K^6,~x\in K^4,~y\in K^3$ with $K$ the 
canonical bundle on Riemann surface parameterized by $z$. The holomorphic 3-form is taken as
\begin{equation}
\Omega= {dx \wedge dy \wedge dz  \over  w};
\end{equation}
This curve is expressed  precisely in the form we found in \eqref{6d}.  In class ${\cal S}$ construction, the punctures are labeled by nilpotent orbits, and the ${\cal N}=2$ geometry is found by calculating the spectral curve of the corresponding Hitchin system. The crucial data is to identify the local pole structures to the various differentials $\epsilon_i(z)$. This data has been worked out in 
\cite{Chacaltana:2014jba} for $E_6$ class S theory, and the result is:
the pole structure of $\epsilon_i$ near the full puncture is $(1,4,5,7,8,11)$ and the order of pole near the minimal puncture  is
$(1,1,2,2,2,3)$ \cite{Chacaltana:2014jba}. Comparing with the pole structure we found above, we see that the 4d theory found above is just $E_6$ theory on a sphere with two full punctures and one minimal puncture, and this proves the conjecture in
 \cite{Ohmori:2015pua}. Here we derive the structure of the punctures directly using geometric engineering of the theory
 in 6d and using mirror symmetry! Similarly for rank $N$ conformal matter setting $f=0$
one obtains the 4d class ${\cal S}[E_6]$ theory corresponding to a sphere with two full punctures and N simple punctures.

 
\subsection{$(E_7,E_7)$ conformal matter}

The LG mirror potential for the geometry corresponding to the geometric engineering of the $(E_7,E_7)$ conformal matter of eqn.\eqref{LGconformatta} can be equivalently rewritten in the form
\begin{equation}
W = x_1^2+x_2^3x_3+x_3^4+y_1^{4N}+y_2^{4N}+\sum f_i(y)g_i(x).
\end{equation}
The deformations are given by the weight one monomials built out of the allowed deformations of $W_{T^2/\Z_4}$, and the allowed $y$ monomials which are of the form $y_1^iy_2^j,~i+j=pN,~p=0,1,2,3,4$. By tuning the parameters we obtain
\be
\begin{cases}
& W = x_1^2+x_2^3 g(x_3, y_1, y_2)+g(x_3, y_1, y_2)^3x_3+g(x_3, y_1, y_2)^4 \nonumber\\
& g(x_3, y_1, y_2)=x_3+y_1^N+y_2^N+\sum_{i=1}^{N-1} y_1^i y_2^{N-i}.
\end{cases}
\ee
There is an $E_7$ singularity over $ g(1, y_1, y_2)=0$ at the affine patch $x_3=1$, and the $g(x_3, y_1, y_2)^4$ term becomes irrelevant. Again, we find a Riemann sphere with $N+2$ punctures described by $ g(1, y_1, y_2)=0$  . 
Similarly, we can prove that the puncture at $y_1=0$ or $y_2=0$ are full punctures, and the puncture at $y_1\rightarrow\infty$ is the simple puncture. The proof goes parallel to the $E_6$ case and we
leave the details for the interested reader and list only the result. The $\cN=2$ geometry (using $y_1$ as the coordinate for the punctured sphere) is 
 \be
 \begin{cases}
 &f = x_1^2+x_2^3\rho+\rho^3+(m_1+m_1^{'}y_1) x_2\rho^2+(m_2+u_1y_1+m_2^{'} y_1^2)\rho^2 + \nonumber\\ 
& ~~(m_3+u_2y_1+m_3^{'} y_1^2) x_2\rho +(m_4+u_3 y_1+m_4^{'} y_1^2)x_2^2 +(m_7+u_4 y_1+u_5y_1^2+m_4^{'} y_1^3)\rho +\nonumber\\
&(m_6+u_6 y_1+u_7y_1^2+m_5^{'} y_1^3)x_2 +(m_7+u_8y_1+u_9y_1^2+u_{10}y_1^3+m_7^{'}y_1^4)=0.   \nonumber\\
&\rho=(1+y_1+y_2).
 \end{cases}
 \ee
Here $m_i$ are the mass parameters, $u_i$ are Coulomb branch vevs, and the holomorphic 3-form is $\Omega={dx_2\wedge d\rho \over x_1}\wedge{d y_1\over y_1}$.
 The invariant polynomial for $E_7$ theory is parameterized by the differentials $(\epsilon_{2}, \epsilon_6, \epsilon_8, \epsilon_{10}, \epsilon_{12}, \epsilon_{14}, \epsilon_{18})$,
 which are the coefficients before the monomials $(\tilde{x_2}\rho^2,\tilde{x_2}^2, \tilde{x_2}\rho, \rho^2, \tilde{x_2}, \rho,1)$.
 The order of poles near the three punctures are :
\begin{align}
& y_1=0~\text{and}~y_2=0:~~~(1,5,7,9,11,13,17),    \nonumber\\
 & y_1=\infty:~~~(1,2,2,2,3,3,4).
\end{align}

The $\cN=2$ geometry for $E_7$ class ${\cal S}$ theory can be written in the following form: 
\begin{align}
& w^2+x^3+xy^3+\epsilon_{2}(z) x^2y+\epsilon_6(z) x^2+ \epsilon_8(z) x y+\epsilon_{10}(z) y^2+  \nonumber\\
& \epsilon_{12}(z) x+\epsilon_{14}(z) y+\epsilon_{18}(z)=0. 
\end{align}
Here $w\in K^9,~x\in K^6,~~y\in K^4$ with $K$ the canonical bundle of the Riemann surface.  The full puncture is labeled using the regular Nilpotent orbit of $E_7$ lie algebra, and its 
order of pole to the differential $\epsilon_i(z)$ is $(1,5,7,9,11,13,17)$, and 
the simple puncture is labeled by the minimal Nilpotent orbit of $E_7$, and it has pole structure $(1,2,2,2,3,3,4)$.\footnote {We thank Oscar Chacaltana for confirming this result based on unpublished work on class ${\cal S}$ $E_7$ theory.}
So the 4d theory we find 
is a class ${\cal S}[E_7]$ on a sphere with two full punctures and one simple puncture. For rank $N$ conformal matter we 
get a 4d class ${\cal S}$ theory defined by the $E_7$ theory on a sphere with two full punctures and N simple punctures.

\subsection{$(E_8,E_8)$ conformal matter}
Tuning the parameters of the LG mirror potential in eqn.\eqref{LGconformatta} we obtain
\be
\begin{cases}
&W =  x_1^2+x_2^3+(g(x_3,y_1,y_2))^5x_3+ g(x_3, y_1, y_2)^6 \nonumber\\
& g(x_3, y_1, y_2)=x_3+y_1^N+y_2^N+\sum_{i=1}^{N-1} y_1^i y_2^{N-i},
\label{e8}
\end{cases}
\ee
and there is an $E_8$ singularity over $ g(1, y_1, y_2)=0$ at the affine patch $x_3=1$. Again, we find a Riemann sphere with $N+2$ punctures. 
Similarly, we can prove that the puncture at $y_1=0$ or $y_2=0$ are full punctures, and the puncture at $y_1\rightarrow \infty$ is the simple puncture. The proof is parallel to the $E_6$ case, and 
the geometry is 
\be
 \begin{cases}
 &f = x_1^2+x_2^3+\rho^5+(m_1+m_1^{'}y_1) x_2\rho^3+(uy_1^2)\rho^4+(m_2+u_1y_1+m_2^{'} y_1^2)x_2\rho^2+ \nonumber\\ 
& (m_3+u_2y_1+u_3y_1^2+m_3^{'} y_1^3) \rho^3 +(m_4+u_4 y_1+u_5y_1^2+m_4^{'} y_1^3) x_2\rho+\nonumber\\
&(m_5+u_6 y_1+u_7y_1^2+u_8y_1^3+m_5^{'} y_1^4) \rho^2+(m_6+u_9y_1+u_{10}y_1^2+u_{11}y_1^3+m_6^{'}y_1^4)x_2+ \nonumber\\
&(m_7+u_{12}y_1+u_{13}y_1^2+u_{14}y_1^3+u_{15}y_1^4+m_7^{'}y_1^5)\rho+  \nonumber\\
&((m_8+u_{16}y_1+u_{17}y_1^2+u_{18}y_1^3+u_{19}y_1^4+u_{20}y_1^5+m_8^{'}y_1^6)=0; \nonumber\\
&\rho=(1+y_1+y_2).
 \end{cases}
 \ee
The invariant polynomial for $E_8$ theory is parameterized by the following differentials on Riemann sphere:
$(\epsilon_{2}, \epsilon_8, \epsilon_{12}, \epsilon_{14}, \epsilon_{18}, \epsilon_{20}, \epsilon_{24},\epsilon_{30})$ which are the coefficients before 
the monomial $(x_1\rho^3,x_2\rho^2,\rho^3,x_2\rho,\rho^2,x_2,\rho,1)$, and 
the order of the poles near three punctures are
\begin{align}
&  y_1=0~\text{and}~y_2=0:~~~(1,7,11,13,17,19,23,29),    \nonumber\\
 &y_1=\infty:~~~(1,2,3,3,4,4,5,6).
\end{align}
Moreover, we find a new term $uy_1^2 \rho^4$ which gives us a dimension 6 operator, and the order of pole of this differential  at the simple puncture is 
$2$, and the order of pole of the  full puncture of this differential is $5$.

The $\cN=2$ geometry for $E_8$ class ${\cal S}$ theory can be written in terms of the following Calabi-Yau geometry: 
\begin{align}
& w^2+x^3+y^5+\epsilon_{2}(z) xy^3+\epsilon_8(z) xy^2+ \epsilon_{12}(z) y^3+ \epsilon_{14}(z) x y+\epsilon_{18}(z) y^2+  \nonumber\\
& \epsilon_{20}(z) x+\epsilon_{24}(z) y+\epsilon_{30}(z)=0. 
\end{align}
Here $w\in K^{15},~x\in K^{10},~~y\in K^{6}$ with $K$ the canonical bundle of the Riemann surface.  The full puncture is labeled using the regular Nilpotent orbit of $E_8$ lie algebra, and its 
order of pole for the differential $\epsilon_i(z)$ is $(1,7,11,13,17,19,23,29)$. The 
simple puncture is labeled by the minimal Nilpotent orbit of $E_8$, and its local contribution to the pole structure is rather subtle, in fact, the basic invariant involves $\epsilon_6$ 
and the order of pole near the simple puncture is $2$, and the order of pole near the other basis differentials is  $(1,2,3,3,4,4,5,6)$ \footnote{We thank Oscar Chacaltana for confirming part of this result based on unpublished 
class ${\cal S}$ theory analysis. }. So the 4d theory we find 
is a class ${\cal S}$ theory defined using 6d $E_8$ $(2,0)$ theory on a sphere with two full punctures and one simple puncture. For rank $N$ conformal matter we 
get a 4d class ${\cal S}$ theory defined by $E_8$ theory on a sphere with two full punctures and N simple punctures.

\subsection{$(G, G^{'})$ conformal matter}
In this section we find the geometry associated with $(G,G')$ conformal matter systems discussed in section 3, for the cases 
 $(E_7,SO(7)$, $(E_8,G_2)$, and $(E_8, F_4)$, which preserved the global symmetries.  We
then find a 4d SCFT by locating the most singular point in moduli space.    We consider an orbifold $(T^2\times \C^2)/ \mathsf{G}$ where the orbifold action is 
\begin{equation}
g_1:~~(z; z_1, z_2)\rightarrow (\alpha z; z_1, \alpha^{-1} z_2),~~\qquad g_2:~~(z; z_1, z_2)\rightarrow (\eta z; \eta^{-1}  z_1, z_2).
\label{orbi}
\end{equation}

\subsubsection{$(E_7,SO(7))$}

Let us take $\alpha=\exp({2\pi i\over 4}),~\eta=\exp({2\pi i\over 2} )$. We expect that this theory describes  $(E_7,SO(7))$ conformal matter due to the orbifold action, see section 3. 
We put this 6d theory on $T^2$ leading to the LG
\be
\begin{aligned}
&W = x_1^2+x_2^4+x_3^3 x_2+ a x_2x_3 x_1+y^2+y_2^{4}+y(x_2 ^2+x_1+x_3^2)+y y_2(x_2+x_3)+\\
& ~~~~y_2(x_2x_1+x_2^3)+y_2^2(x_2 ^2+x_1+x_3^2)+y_2^3(x_2 +x_3)+y_1^2y_2^2.
\end{aligned}
\ee 
Here we suppress the coefficients before each allowed deformation. There are a total of 14 parameters which agrees with the result from 6d tensor branch description.  We have used a different but equivalent curve for $T^2/Z_4$ part, 
\begin{equation}
W=x_1^2+x_2^4+x_3^3 x_2+x_2x_3 x_1, 
\end{equation}
from which we have the relation
\begin{equation}
x_1=x_2 x_3,~~x_2^3=x_3^3+x_1x_3,~~x_3^2 x_2= x_2 x_1;
\end{equation}
Here we ignore the unimportant numerical factors, and the ring is generated by these generators $( x_2, x_3, x_1, x_2^2, x_3^2, x_1x_2,x_2^3,,1)$.
We can tune the parameter so that $W$ becomes:
\begin{equation}
W=(x_1+y+y_2^2)^2+y_2 x_2^3+x_3^3 x_2+x_2^4+x_1x_2x_3
\end{equation}
This potential leads to a singularity at $x_1+y+y_2^2=0,~x_2=0,~x_3=0$, and the singularity type is $E_7$ (we can absorb $y_2$ by redefining the coordinates as $y_2\neq 0$). So
there is an $E_7$ singularity over the curve defined by $1+y+y_2^2=0$ by  going to affine patch $x_1=1$ (note that $x_2^4$ term is irrelevant). There are three punctures 
at $y=0,y_2=0$ and $y_2=\infty$.   

We now analyze the pole structure near various punctures. Our geometry has the following form: 
\be
\begin{cases}
& \rho^2+y_2x_2^3+x_3^3x_2+(y_2)x_2x_3^2+(y+y_2^2)x_2^2+(y+y_2^2)x_2x_3+(y+y_2^2)x_3^2+(y_2y+y_2^3)x_2 \nonumber\\
&+(y_2y+y_2^3)x_3+(y_2^4+y_2^2y+y^2)=0 \nonumber\\
& \rho=1+y+y_2^2
\end{cases} 
\ee
Following the same analysis as we have done for the  conformal matter, we find that $y=0$ is a $E_7$ full puncture,
and $y_2\rightarrow \infty$ is a $E_7$ simple puncture. The puncture near $y_2\rightarrow 0$ is not a full puncture. To get the correct flavor symmetry, the puncture has to be of $(A_3+A_1)$ type \cite{Chacaltana:2012zy}, 
as the order of pole structure and constraint for this puncture is not available yet, we could not 
compare our result with class ${\cal S}$ construction.  The central charges for this 4d theory can be computed using the methods of \cite{Chacaltana:2012zy}, and we obtain: 
\begin{equation}
a={385\over24}~~c={119\over6}.
\end{equation}
By analyzing this particular class $\cS$ theory, we find that it is a combination of an interacting 
SCFT and some free hypers. This interacting SCFT has three Coulomb branch operators with scaling dimensions 6,8 and 12. 
Such model can also be realized by using $E_6$ class $\cS$ theory with a full puncture, a simple puncture and a $2A_1$ puncture. 

Notice that the maximal singular point we find is slightly different from the one suggested in \cite{Ohmori:2015pua} for $(E_7, SO(7))$ conformal matter. In 
that paper, they identified the irreducible part, namely an $E_6$  class ${\cal S}$ theory. This theory has 
central charge $a={119\over 8},~c={35\over 2}$.  Comparing with the central charge of the $E_7$ version, we find a difference $\delta a={7\over 6}$ and 
$\delta c={7\over 3}$, and this is consistent with the interpretation that there are 28 decoupled free hypers transforming in the  (${1\over 2} \bf{56}$) of $E_7$ for the class $\cS[E_7]$ realization. 

We can also directly locate the $E_6$ version of this theory from the geometry we have obtained.   Using an equivalent form for mirror of $T^2/Z_4$ (replacing $x_3^3 x_2$ term with $x_2^4$ term) the singular LG is 
\begin{equation}
W=(x_1+y+y_2^2)^2+y_2 x_2^3+x_3^4+x_2^4+x_1x_2x_3.
\end{equation}
This geometry has an $E_6$ singularity over $1+y+y_2^2=0$, and the $x_2^4$ term is irrelevant. The full $\mathcal{N}=2$ geometry is 
\be
\begin{cases}
& \rho^2+y_2x_2^3+x_3^4+(y_2)x_2x_3^2+(y+y_2^2)x_2x_3+(y+y_2^2)x_3^2+(y_2y+y_2^3)x_2 \nonumber\\
&+(y_2y+y_2^3)x_3+(y_2^4+y_2^2y+y^2)=0.  \nonumber\\
& \rho=1+y+y_2^2;
\end{cases} 
\ee
By analyzing the deformations and 
pole structure 
we find an $E_6$ full puncture, an $E_6$ minimal puncture, and a $2 A_1$ puncture, which is exactly the one 
suggested in \cite{Ohmori:2015pua}.   This also gives a 6d explanation of the enhancement of the
global symmetry for this theory from $E_6\times SO(7)\times U(1)\rightarrow E_7\times SO(7)$ \cite{Chacaltana:2014jba} .

\subsubsection{$(E_8, F_4)$}

Next consider the same type of orbifold (\ref{orbi}) with $\alpha=\exp({2\pi i\over 6}),~\eta=\exp({2\pi i\over 3} )$. The LG mirror potential is
\begin{align}
&W = x_1^2+x_2^3+x_3^6+a x_1 x_2x_3+y_1^{6}+y_2^{6}+y(x_2 x_3^2+x_3^4)+y^2 (x_2+x_3^2)+ \nonumber\\
& ~~y_2(x_2x_3^3)+y_2^2(x_2x_3^2+x_3^4)+y_2^3(x_2x_3 +x_3^3)+y_2^4(x_2+x_3^2)+y_2^5(x_3)+ \nonumber\\
& ~~y y_2(x_2x_3 +x_3^3)+yy_2^2(x_2+x_3^2)+yy_2^3(x_3)+yy_2^4+ \nonumber\\
& ~~y^2 y_2(x_3)+y^2y_2^2
\end{align}
There are a total of 21 parameters which match with the tensor branch description of $(E_8, F_4)$ conformal matter. We can tune the parameters in such a way that
\begin{equation}
W = (x_1+y+y_2^2)^2y_2^2+x_2^3+y_2 x_3^5
\end{equation}
So at the affine patch $x_1=1$ there is a $E_8$ singularity over the curve $1+y+y_2^2=0$, which defines 
a three punctured sphere. By analyzing the pole structure, we find an $E_8$ full puncture and a simple puncture. To match
the flavor symmetry, the third puncture has to be a $D_4$ puncture, and the pole structure of this puncture is not available yet so we could not compare our result
with the class ${\cal S}$ construction.

\subsubsection{$(E_8,G_2)$}

Finally, consider the same type of orbifold (\ref{orbi}) with $\alpha=\exp({2\pi i\over 6}),~\eta=\exp({2\pi i\over 2} )$.  We have
\begin{align}
&W = x_1^2+x_2^3+x_3^6+a x_1x_2x_3+y^{3}+y_2^{6}+y(x_2x_3 +x_3^3)+ \nonumber\\
& y_2(x_2x_3^3)+y_2^2(x_2x_3^2+x_3^4)+y_2^3(x_2x_3 +x_3^3)+y_2^4(x_2+x_3^2)+y_2^5(x_3)+ \nonumber\\
& y_1^3 y_2(x_2+x_3^2) +y_1^3y_2^2(x_3)+yy_2^3.
\end{align}
There are a total of 15 parameters which matches the result from tensor branch of $(E_8,G_2)$ theory. The singular deformation gives
\begin{equation}
W = (x_1+y+y_2^3)^2+x_2^3+y_2 x_3^5,
\end{equation}
so again at the affine patch $x_1=1$ we get a $E_8$ singularity over the curve $(1+y+y_2^3)=0$, and 
we find a three punctured sphere. By analyzing the 
puncture type, we find an $E_8$ full puncture and an $E_8$ minimal puncture. To get the 
correct flavor symmetry, the other puncture has to be $E_6(a_3)$ puncture. The pole structure of this puncture is not available yet to compare with our result. Notice that the number of parameters in 4d is larger than the naive count from the 6d tensor branch description, this is expected from our discussion in section \ref{comicbook}.

\subsection{Other examples}

For the 6d minimal conformal matter models of type $(G,G)$ compactified on $T^2$ we have discussed one example which is not in class $\cS$, and one which is in class $\cS$ at genus 0. Clearly, along the lines of what we have observed in section \ref{comicbook} we expect to find even more inequivalent 4d limits as class $\cS$ theories. The purpose of this section is to provide an example of such sort starting from the mirror geometry.  We will not attempt
to find all such inequivalent 4d limits, but simply provide an existence proof of higher genus versions of class $\cS$ theories coming from 6d
conformal matter of $E$-type.

For concreteness consider the rank $N$ $E_8$ conformal matter whose curve is given in ($\ref{e8}$).  We know that this can give either a 4d theory of class $\cS$ $E_8$ $(2,0)$ theory on a sphere with 2 full punctures and $N$ simple punctures as we discussed in last subsection, or it can give
an affine $E_8$  quiver gauge theory with middle gauge group $SU(6N)$ by going to the $A_{6N-1}$ singularity locus.
We can find other limits as well, for example,  we can
tune parameters in ($\ref{e8}$) to get the following singular geometry:
\begin{equation}
W=x_1^2+x_2^3+(x_3^2+f_{2N}(y_1,y_2))^3,
\end{equation}
which corresponds to a $D_4$ theory of class $\cS$ on the hyper-elliptic curve
$x_3^2+f_{2N}(y_1,1)=0$, of genus $g=N-1$ with $4+4N$ punctures of some type which can be determined using the same methods of the last subsection.

\section{Conclusion}

In this paper we have initiated a systematic study of the toroidal compactification of $6d_{(1,0)} \to 4d_{(\cN=2)}$ based on
geometric engineering of these theories and employing mirror technology to solve for the effective 4d vacuum geometry.   Along the way we have established a simple dictionary between those 6d (1,0) SCFTs which can be realized as $F$-theory on orbifolds and their LG mirrors. The details of dictionary have been spelled out for those models which are abelian orbifolds, but we believe that along these lines it should be possible to analyze also all the other models of this sort. We have found that the map from a given 6d (1,0) SCFT to 4d is far from 1-to-1.  We showed
this both for the $A$-type 6d $(1,0)$ theories as well as the orbifold $(1,0)$ theories.    We identified several possibilities which are allowed. One of our findings along this analysis is that there are several possibilities which are mutually exclusive: this is nicely exemplified by means of the 4d SCFT associated to conformal matters of type $(G,G)$. On one hand we have found toroidal reductions which admitted an exact $SL(2,\Z)$ action, but have broken the flavor symmetry, on the other we have found examples which are in class $\cS$ and have large flavor symmetry, but the exact $SL(2,\Z)$ is
sacrificed. We have also found a 6d explanation of why the moduli of affine ${\cal N}=2$ ADE quivers is flat ADE connections on $T^2$.

An interesting result we found is that the curve which one wraps the (2,0) theory onto for a class $\cS$ engineering, emerges spontanously from the mirror geometry of the $T^2$ compactification of $(1,0)$ theories. Moreover, we have also discussed how, starting from the mirror, one can read off the puncture data of class $\cS$ with very little effort.   Our findings point towards the possibility of classifying
all ${\cal N}=2$ theories in 4d by simply studying quasi-homogenous polynomials which have $\hat c<2$ and
if they have a curve singularity, having $\hat c <1$ singularity along the curve. 

Let us also mention that in this project our focus has been the fate of the local structure of the 6d SCFT upon compactification. It would be  interesting to also study the fate of the surface defects of the 6d (1,0) theory upon compactification, perhaps along the lines suggested in\cite{DelZotto:2015isa}.

Finally the most natural next step is to study compactifications of $(1,0)$ theories on Riemann surfaces, and obtain ${\cal N}=1$ theories
in 4d.  Examples of this type have been studied recently in \cite{Gaiotto:2015usa,Franco:2015jna}, and at the level of holography in \cite{Apruzzi:2015wna,Apruzzi:2015zna,Karndumri:2015rsa}.

\section*{Acknowledgements}

The work of MDZ and CV is supported by NSF grant PHY-1067976.  The work of DX is supported by Center for Mathematical Sciences and applications at Harvard University.  We would like to thank O. Chacaltana, D. Freed, B. Haghighat, J. Heckman, G. Lockhart and T. Rudelius for valuable discussions.


\appendix

\section{$D_4$ AD point for $\cO(-3)$ on $T^2$}\label{D4singularata}
Let us start with the LG model
\be\label{LG3333333}
W=\frac{x_1^3 + x_2^3+x_3^3+y_1^3 + y_2^3}{3} + \alpha \, x_1 x_2 x_3 + y_1 y_2 \sum_i \beta_i x_i,
\ee
The Jacobian ideal of such model is
\be\label{EOMz3333}
\begin{cases}
x_i^2 + \alpha\, x_j x_k + \beta_i y_1 y_2 = 0,\\
y_i^2 + y_j \sum_i \beta_i x_i = 0. 
\end{cases}
\ee
Let us set $\beta_2 = \beta_3 = 0$ and keep $\beta_1 = \beta \neq 0$. We have that \eqref{EOMz3333} entails
\be
\frac{y_1^2}{y_2} = \beta x_1 = \frac{y_2^2}{y_1}, \qquad \left(\frac{y^2_1}{\beta y_2}\right)^2 +\beta y_1 y_2 = -\alpha x_2 x_3.
\ee
Choosing $y_1 = y_2 = 1$ we obtain $x_1 = 1/\beta$ from the first equation, while the second gives
\be
\frac{1}{\beta^2} +\beta = -\alpha x_2 x_3
\ee
Plugging in the values $y_1=y_2 =1$, $\beta_2 = \beta_3 = 0$, $\beta=\beta_1$ into the equation $W=0$ where $W$ is in eqn.\eqref{LG3333333}, one obtains
\be
x_2^3 + x_3^3 + P_3(1/\beta) =0, \qquad P_3(0) = 2/3.
\ee
where $P_3$ is a polynomial of degree 3 in $1/\beta$. Clearly we can tune $1/\beta$ to a root of $P_3$, which gives the desired singularity at $x_2 = x_3 =0$, $y_1=y_2=1$, $x_1 = 1/\beta$.

\section{Properties of 4d $(E^{(1,1)}_n, SU(N))$ theories}\label{ellipticSCFTs}
So we have located 4d SCFTs whose ${\cal N}=2$ geometry has the following form:
\begin{align}
& (E_6^{(1,1)},SU(N)):~~x_1^3+x_2^3+x_3^3+w^N=0; \nonumber\\
& (E_7^{(1,1)},SU(N)):~~ x_1^2+x_2^4+x_3^4+w^N=0; \nonumber\\
& (E_8^{(1,1)},SU(N)):~~  x_1^2+x_2^3+x_3^6+w^N=0.
\end{align}
We  use $(E_n^{(1,1)}, SU(N))$ to label them as the corresponding BPS quiver is the product of double affine $E$ type quiver and 
the  SU(N) Dynkin quiver. These models are examples of the $E_n^{(1,1)}\circledast G$ systems constructed in \cite{Cecotti:2013lda} where this other notation was used to emphasize that the product is not a standard product in between quivers, because the elliptic quivers have non-trivial potential. 

Using the above singular curve, one can read the spectrum of operators parametrizing the Coulomb branch of these models. We obtain:

\begin{center}
  \begin{tabular}{|l|c|c|c|c|c|r| }
    \hline
      ~& Coulomb & mass &marginal &Relevant &BPS quiver&parameter\\ \hline
       $(D_4^{1,1}, SU(2k))$&6k-5&4&5&0&12k-6& 6k+4\\ \hline
       $(D_4^{1,1}, SU(2k+1))$&6k&0&1&4&12k&6k+5 \\ 
 \hline
  \end{tabular}
\end{center}

\begin{center}
  \begin{tabular}{|l|c|c|c|c|c|r| }
    \hline
      ~& Coulomb & mass &marginal &Relevant &BPS quiver&parameter\\ \hline
       $(E_6^{1,1}, SU(3k))$&12k-7&6&7&0&24k-8& 12k+6\\ \hline
       $(E_6^{1,1}, SU(3k+1))$&12k&0&1&6&24k&12k+7 \\ \hline
       $(E_6^{1,1}, SU(3k+2))$&12k+6&0&1&6&24k+8&12k+13 \\
 \hline
  \end{tabular}
\end{center}

\begin{center}
  \begin{tabular}{|l|c|c|c|c|c|r| }
    \hline
      ~& Coulomb & mass &marginal &Relevant &BPS quiver&parameter\\ \hline
       $(E_7^{1,1}, SU(4k))$&18k-8&7&8&0&36k-9& 18k+7\\ \hline
       $(E_7^{1,1}, SU(4k+1))$&18k&0&1&7&36k&18k+8 \\ \hline
          $(E_7^{1,1}, SU(4k+2))$&18k+3&3&4&4&36k+9&18k+14 \\ \hline
       $(E_7^{1,1}, SU(4k+3))$&18k+9&0&1&7&36k+18&18k+17 \\
 \hline
  \end{tabular}
\end{center}

\begin{center}
  \begin{tabular}{|l|c|c|c|c|c||r| }
    \hline
      ~& Coulomb & mass &marginal &Relevant &BPS quiver&parameter\\ \hline
       $(E_8^{1,1}, SU(6k))$&30k-9&8&9&0&60k-10&30k+8 \\ \hline
    $(E_8^{1,1}, SU(6k+1))$&30k&0&1&8&60k&30k+9 \\ \hline
     $(E_8^{1,1}, SU(6k+2))$&30k+4&2&3&6&60k+10&30k+15 \\ \hline 
     $(E_8^{1,1}, SU(6k+3))$&30k+8&4&5&4&60k+20& 30k+21\\ \hline
      $(E_8^{1,1}, SU(6k+4))$&30k+14&2&3&6&60k+30&30k+25 \\ \hline
       $(E_8^{1,1}, SU(6k+5))$&30k+20&0&1&8&60k+40& 30k+29\\ 
        \hline
  \end{tabular}
\end{center}

Since these theories all have exactly marginal deformations, we would like to find a weakly coupled gauge theory description: this is precisely how these models have been introduced in \cite{Cecotti:2013lda}. In such $S$-duality frame these systems have the following form: 
\be
\begin{aligned}
&( D_4^{1,1},G):~~~G-D_2(G)\oplus D_2(G)\oplus D_2(G)\oplus D_2(G),\\
&( E_6^{1,1},G):~~~G-D_3(G)\oplus D_3(G)\oplus D_3(G),\\
&( E_7^{1,1},G):~~~G-D_2(G)\oplus D_4(G)\oplus D_4(G),\\
&(E_8^{1,1},G):~~~G-D_2(G)\oplus D_3(G)\oplus D_6(G).
\label{gauge}
\end{aligned}
\ee
here $D_p(G)$ denotes an Argyres-Douglas type theory with non-abelian $G$ flavor symmetry. Notice that not all of them are theories with a single gauge group, as the Argyres-Douglas matter might have gauge group factors (This is the case 
if there is a dimension two operator in the spectrum). 

This structure has the exact same form as predicted from orbifold geometry as discussed in the main body of the text.

Taking seriously the geometric realization of the decoupling limit discussed in \cite{Xie:2012hs,Cecotti:2012jx} we provide a type IIB description for the $D_p(G)$ theories. We claim that the Coulomb branches of these models are characterized by the geometries
\begin{equation}
0 = e^{-px}+W_G(y,z,w) + \text{deformations}
\end{equation}
where $W_G(y,z,w)$ is the standard polynomial of ADE singularity. Notice that these are not isolated singularities, but, from the findings in the main body of the text, this is indeed allowed, as long as the geometry has a scaling symmetry and no scales in it, to identify it with the IIB description of a SCFT. Let's study $G=SU(N)$ in detail, then the curve is 
\begin{equation}\label{dpgeometr}
e^{-px}+y^2+z^2+w^N=0,
\end{equation}
Let's now review the computation of the spectrum, which is given by the coefficients before the monomials $e^{-lx}w^a,~~0\leq l<{p-1},~~0\leq a \leq N-2$. The scaling dimension of the coefficient before 
the deformation is 
\begin{equation}
[u_{la}]={2(1-Q_{la})\over (2-\hat{c})}={(Np-ap-lN)\over p}.
\end{equation}
Notice that we have operators with dimension $(N, N-1,\ldots, 2)$ from the monomial 
$w^a,~~0\leq a\leq N-2$, and they should be interpreted as the mass parameters for $SU(N)$ flavor symmetry. 
\begin{center}
  \begin{tabular}{|l|c|c|c|c|c|r| }
    \hline
      ~&~  & Coulomb & mass &marginal &Relevant &BPS quiver\\ \hline
   $D_2(SU(N))$ &N=2k &k-1& 1+(N-1)&1&0&2(N-1) \\ \hline
~ &N=2k+1 &k& 0+(N-1)&0&1&2(N-1)\\ \hline
      $D_3(SU(N))$ &N=3k &3k-2& 2+(N-1)&2&0&3(N-1) \\ \hline
~ &N=3k+1 &3k& 0+(N-1)&0&2 &3(N-1)\\ \hline
~ &N=3k+2 &3k+1& 0+(N-1)&0&2 &3(N-1)\\ \hline
        $D_4(SU(N))$ &N=4k &6k-3& 3+N-1&3&0&4(N-1) \\ \hline
~ &N=4k+1 &6k& 0+(N-1)&0&3& 4(N-1)\\ \hline
~ &N=4k+2 &6k+1& 1+(N-1)&1&2&4(N-1)\\ \hline
 ~ &N=4k+3 &6k+3& 0+(N-1)&0&3&4(N-1)\\ \hline
   
     $D_6(SU(N))$ &N=6k &15k-5&5+ N-1&5&0&6(N-1) \\ \hline
~ &N=6k+1 &15k& 0+(N-1)&0&5&6(N-1)\\ \hline
~ &N=6k+2 &15k+2& 1+(N-1)&1&4&6(N-1)\\ \hline
~ &N=6k+3 &15k+4& 2+(N-1)&2&3&6(N-1)\\ \hline
~ &N=6k+4 &15k+7& 1+(N-1)&1&4&6(N-1)\\ \hline
~ &N=6k+5 &15k+10& 0+(N-1)&0&5&6(N-1)\\ \hline

 \hline
  \end{tabular}
\end{center}

There are some further properties of $D_p(SU(N)$ theory:
\begin{itemize}
\item If we gauge $G$ flavor symmetry of the theory, its contribution to $\beta$ function is 
\begin{equation}
D_p(SU(N))=N\,{p-1\over p}
\end{equation}
\item The following list is Lagrangian:
\be
D_p(SU(pm):~~SU(m)-SU(2m)-SU(3m)-...-SU((p-1)m)-mp
\ee
\item If $\text{gcd}(p,N)\neq 1$, then there is an exact marginal deformation, and the theory can be written as a gauge theory coupled to Argyres-Douglas matter. 
\end{itemize}

Using the above information of $D_p(G)$ theory, and the gauging patter listed in \ref{gauge}, one can check that the spectrum 
from weakly coupled gauge theory is the same as the $(E_n^{1,1},G)$ theory studied in last subsection. 

 Using the above properties of $D_p(G)$ theory, we find that $(E_6^{1,1}, SU(3m))$, $(E_7^{1,1}, SU(4m))$ and $(E_8^{1,1}, SU(6m))$ 
 are actually the affine quiver gauge theory of $E_n$ shape, see figure \ref{affine}. The gauge coupling of the middle quiver is identified with the complex structure of the torus. The gauge couplings of other quiver nodes are shown to be governed by the 
moduli space of $E_n$ type flat connections on $T_2$. It is very suggestive that our construction for these affine quiver gauge theories involve a $T^2$ and $E$ type gauge algebra in six dimension, and it is natural that moduli space of $E_n$ flat connection on $T^2$ appears.

\section{Non-Higgsable models on $T^2$ and $(E_n^{(1,1)},G)$ theories} 
In this appendix we consider the abelian orbifolds discussed in section \S.\ref{subsaborbfs}. These are non-Higgsable theories of $A$--type. The corresponding LG mirrors are completely determined by the data $(p,q,k)$ which can be read off from table \ref{genaborb}. With the same notation as above
\be\label{LGaborbifoldi}
W = W_{T^2/\Z_r}(x_i) + y_1^{p} +y_2^{p} + \sum_{\ell=0}^k\sum_{j=1}^{m_\ell} \sum_a \, t_{\ell,j,a} \, \varphi_{j,\ell/k}(x_i) \, y_1^{a} \, y_2^{ [a \, q]_p },
\ee
where the sum over $a$ is taken conditionally on $\ell$ only for the values of $a$ which solve the equation
\be
\frac{(a+ [q \, a]_p)}{p}+\frac{\ell}{k} = 1  \qquad\qquad 0\leq a < p
\ee
and the notation $[x]_p$ stands for $x \text{ mod }p$. Notice that we can always proceed as in the previous example and tune such $W$ in such a way that it reduces to
\be
W = W_{T^2/\Z_r}(x_i) + y_1^{p} +y_2^{p} + (y_1^k + y_2^k)^{p/k}
\ee
Then, by proceeding as in the previous example, we obtain an isolated singularity of the type
\be
0 = W_{T^2/\Z_k}(x_i)  + w^{p/k}
\ee 
which corresponds to the $(E_k^{(1,1)} , SU(p/k))$ SCFT, which confirms our prediction based on the M-theory geometry, discussed in section \ref{mtheoryengine}. Looking at table \ref{genaborb} we see that there are some models for which $k$ is a divisor of $p/k$, e.g. $(3,A_N,3)$, $(2,2,2,2,4,2,2,2,2)$, or $(2,4,2)$ which gives respectively $(D_4^{(1,1)} , SU(2N+4))$, $(E_8^{(1,1)},SU(30))$, and $(E_6^{(1,1)}, SU(6)).$ These are lagrangian SCFTs of affine type. If $k$ does not divide $p/k$, we obtain an AD point which always contains some non--lagrangian strongly coupled subsectors. Let us notice that for these families of models we have started in 6d with a theory which was non-Higgsable, we have reduced it on $T^2$ and we have located along its 4d moduli space a theory which has a Higgs branch. This phenomenon is reminiscent of the findings of \cite{Argyres:2012fu}. Understanding its physics is, however, beyond the scope of the present note, and we leave this for future work.


\bibliographystyle{utphys}

\bibliography{downonT2b}

\end{document}